# Synthetic dimensions using ultrafast free electrons

Yiming Pan[1], Bin Zhang[2], Daniel Podolsky[1]

1. Department of Physics, Technion, Haifa 3200003, Israel

2. Department of Electrical Engineering Physical Electronics, Tel Aviv University, Ramat Aviv 6997801, Israel

**Abstract**

We propose a synthetic dimension for ultrafast free electrons based on the discrete energy sidebands in photon-induced near-field electron microscopy (PINEM). The PINEM synthetic dimension can offer a powerful way to coherently shape or modulate free-electron wavefunctions in systems such as ultrafast transmission electron microscopes, dielectric laser accelerators, and quantum free-electron lasers. As examples of this paradigm, we demonstrate Bloch oscillations, diffraction management, and linear optics effects such as Talbot self-imaging in the PINEM lattice. These effects show the potential of PINEM synthetic dimensions as a novel quantum simulation platform.

Photon-induced near-field electron microscopy (PINEM)[1–3] is an ultrafast electron microscopy and spectroscopy approach. It exploits the enhancement of electron-photon interactions via nanostructures illuminated by a laser pulse to study evanescent electromagnetic fields of nanostructures, and is capable of resolving these near-fields both spatially on the atomic scale and temporally in femtosecond timescale. PINEM is a physics of strong-field light-matter interaction that involves the discretized multiphoton emission and absorption processes that occur when an electron pulse passes through the optically stimulated near-fields[1,4–9]. The scattering of the free electron and light gives rise to discrete spectral sidebands equally spaced by light quanta ($\hbar\omega_L$)[3,4], characterized by electron energy loss/gain spectra (EELS/EEGS)[10]. These PINEM sidebands develop as a result of the optical phase-modulation of the ultrafast free-electron wavefunction, opening up new avenues for imaging ultrafast dynamics and plasmonics[1,5,10,11]. As a result, these ultrafast electrons can be coherently manipulated in a variety of controllable ways, allowing for four-dimensional attosecond electron imaging[1,12,13], attosecond electron pulse generation[13,14], Ramsey-type and holographic phase control[15,16], quantum state reconstruction [14] and free-electron qubits[17–19]. Nowadays, PINEM electrons have progressed beyond their initial promise. PINEM interactions, for example, can be employed to create quantum entanglement between electrons and photons[20], resonant interactions with bounded electrons[21–23], anomalous PINEM[24], and to construct a quantum-classical correspondence to dielectric laser accelerators (DLA)[25–28].

Here, we propose a novel direction for exploring synthetic dimensions using PINEM free electrons. The concept of synthetic dimensions has been widely investigated in several proposals with atoms[29–32] and photons[33–37]. A synthetic dimension is an extra dimension used for quantum simulation, enabling a low-dimensional system to simulate high-dimensional dynamics. As a result, synthetic dimensions have received considerable interest in many disciplines of physics[31,34,38]. However, to date, little has been reported on employing PINEM electrons to perform quantum simulations[14,17–19,39].

The basic idea behind PINEM synthetic dimensions exploits a close analogy between laser-modulated electrons in PINEM and the physics of an electron subject to the periodic potential in a crystal. The PINEM synthetic lattice describes the multiphoton processes along the energy axis, in which the absorption or emission of a photon corresponds to hopping processes in the lattice. Indeed, multi-level Rabi oscillations using ultrafast electrons have been experimentally observed by Feist *et al.*[4]. Due to the analogy to quantum walks of light in waveguides[39,40], they suggested that the PINEM electrons may be harnessed as a new type of "quantum hardware"[14]. However, to genuinely simulate quantum-mechanical effects such as Bloch oscillations[41,42] and Wannier-Stark ladders[43] requires the capability to control the hopping magnitudes and phases between neighboring sidebands, as well as the on-site potential of each sideband. Unfortunately, the inelastic scattering in the vicinity of an optically excited metallic nanotip cannot provide such flexibility[4,9].

In this article, we will demonstrate that nanomaterials with periodic structures can offer the flexibility needed to construct the PINEM synthetic dimensions. We find that the PINEM interaction can be mapped into a tight-binding model subject to an effective electric field, whose strength is given by the phase mismatch between the optical cycle and the grating period. Consequently, the effective constant force can lead to Bloch oscillations in the PINEM lattice. Furthermore, we demonstrate how to accelerate PINEM electrons based on the Bloch oscillations[33,44] . In addition, we show the capability of the PINEM platform to mimic linear optics effects by demonstrating diffraction management[45,46], Talbot effect, and negative refraction in the synthetic lattice.

**Setup of free-electron synthetic dimensions** – The PINEM setup for ultrafast electrons is depicted schematically in Fig. 1a. A single electron pulse is ejected from a photoemission gun that has been excited by an ultraviolet femtosecond pulse. Then the electron enters the interaction region of a grating and interacts with the optical near-fields or plasmonic fields triggered by an IR light pulse. The PINEM process may be represented by a relativistically-modified Schrödinger equation of electrons in the presence of the electromagnetic field,

$$i\hbar \frac{\partial}{\partial t} \psi(z,t) = (H_0 + H_I)\psi(z,t). \tag{1}$$

The free-electron Hamiltonian for one-dimensional relativistic dynamics is $H_0 = \varepsilon_0 + v_0(p - p_0) + \frac{(p-p_0)^2}{2\gamma^3 m}$, which is derived by expanding the Dirac equation about the initial momentum $p_0 = \gamma m v_0$ when the spin index is ignored. In a typical PINEM setup, we choose the initial electron kinetic energy $\varepsilon_0 = (\gamma - 1)mc^2 = 200 \ keV$, corresponding to the electron velocity $v_0 = \beta c$ with the relative speed $\beta = 0.7$ and the Lorentz factor $\gamma = 1/\sqrt{1-\beta^2} = 1.4$. The near-field interaction is $H_I = -\frac{e}{2\gamma m}(A \cdot p + p \cdot A)$. Only the longitudinal component of the near-field vector potential impacts ultrafast electron dynamics in the propagation direction (z). In our case, the transverse field components are ignored. We take a realistic nanograting suggested in the setup of dielectric laser accelerators[27], whose longitudinal electric field is given by $E(z,t) = -\frac{\partial}{\partial t}A(z,t)$, where the vector potential is $A(z,t) = A_0 \sin(\omega_L t - qz + \phi_0)$, with $A_0 = -E_0/\omega_L$, electric field strength $E_0$, laser frequency $\omega_L$, wavevector $q = 2\pi/\Lambda$ for a grating with a period $\Lambda$, and the phase delay $\phi_0$.

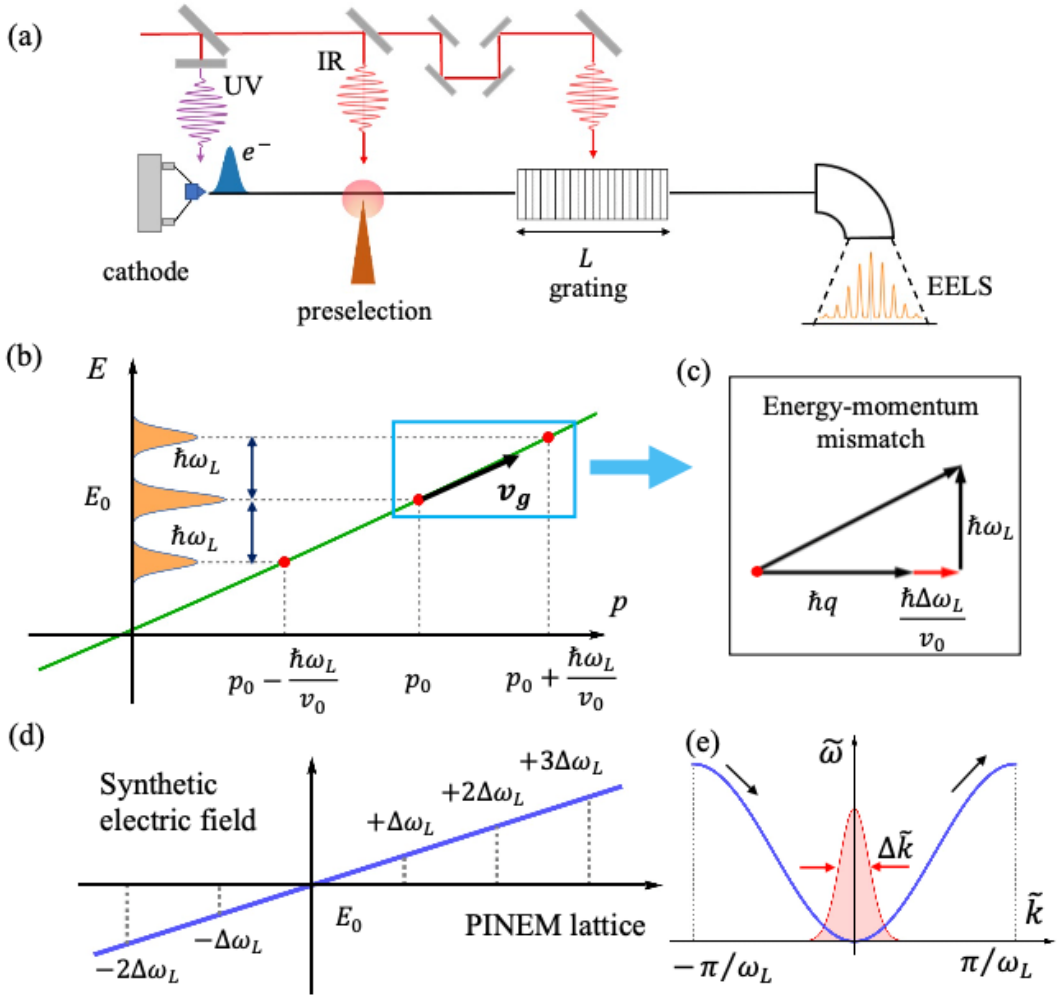

**Fig.1: Schematic diagram of PINEM synthetic dimension using ultrafast free electrons and the induced artificial gauge field in the PINEM lattice.** (a). Ultrafast electron setup: a

photocathode gun produces an electron, which is pre-selected by a tip, passes through a grating and is detected by a magnetic spectrometer. (b) the PINEM sidebands develop when the electron experiences the laser-induced phase modulation on a grating. The sidebands act as a synthetic lattice. A purposeful detuning $\Delta\omega_L$ between the grating period $\Lambda$ and the laser frequency $\omega_L$ results in an effective constant force (d). The phase mismatch for each PINEM lattice is seen in (c). (e) The synthetic dispersion relation represents electron hopping between the PINEM lattice. As a result, the constant force generates a synthetic Bloch oscillation on the PINEM lattice.

The electron wavefunction can be expanded in the Floquet-Bloch basis,

$$\psi(z,t) = \sum_n a_n(t)\, e^{-i(\omega_n t - k_n z)}, \tag{2}$$

with the frequency $\omega_n = \omega_0 + n\omega_L$ and the wave vector $k_n = k_0 + nq$, and $\omega_0 = \varepsilon_0/\hbar, k_0 = p_0/\hbar$, and $n$ is an integer reflecting the net number of photons absorbed by the electron. Note that $a_n(t)$ is a slow-varying envelope in the co-moving frame. Substituting Eq. (2) into the Schrödinger equation (1), we obtain the coupled-mode equations for the amplitude $a_n(t)$, given by $i\frac{\partial}{\partial t} a_n = -n\Delta\omega_L\, a_n + \frac{eA_0}{2\gamma m}\left(\left(k_{n+1} - \frac{q}{2}\right)a_{n+1}e^{i\left(\phi_0 + \frac{\pi}{2}\right)} + \left(k_{n-1} + \frac{q}{2}\right)a_{n-1}e^{-i\left(\phi_0 + \frac{\pi}{2}\right)}\right)$. Here, we defined the frequency detuning $\Delta\omega_L = \left(\omega_L - v_0 q - \frac{n\hbar q^2}{2\gamma^3 m}\right)$. We approximate $\Delta\omega_L \simeq \omega_L - v_0 q$ and $k_{n+1} - \frac{q}{2} \approx k_{n-1} + \frac{q}{2} \approx k_0$ since $k_0 \gg nq$ for ultrafast electrons and $\frac{n\hbar q^2}{2\gamma^3 m} \approx 0$ for small n. Hence, the coupled-mode equations can be simplified as

$$i\frac{\partial}{\partial t} a_n = -n\Delta\omega_L a_n + \kappa a_{n+1} + \kappa^* a_{n-1}, \tag{3}$$

where the hopping amplitude between neighboring PINEM sidebands is $\kappa = \frac{ek_0 A_0}{2\gamma m} e^{i\left(\phi_0 + \frac{\pi}{2}\right)}$. As illustrated in Fig. 1c, the linear dependency of the on-site potential in Eq. (3) acts as the synthetic electric field in the PINEM lattice with an effective force given by $\Delta\omega_L/\omega_L$. The same form as (3) was found previously in a closely-related setup[44]. The tight-binding Hamiltonian is then

$$H = -\sum_n (n\Delta\omega_L) a_n^\dagger a_n + \sum_n \kappa a_n^\dagger a_{n+1} + h.c., \tag{4}$$

Here, $a_n^\dagger, a_n$ are creation and annihilation operators for fermions (i.e., electrons), with the anticommutation relation: $\{a_n^\dagger, a_m\} = \delta_{nm}$. The analytical solutions of Wannier-Stark ladders for PINEM electrons from (3) are presented in the SM file.

**Synthetic band spectrum** - To obtain the exact solution of (3), we perform a gauge transformation $u_n(t) = e^{-in\Delta\omega_L t} a_n(t)$, yielding a simplified coupled-mode equation $i\frac{\partial}{\partial t} u_n(t) = \kappa_1 u_{n+1}(t) + \kappa_1^* u_{n-1}(t)$, with $\kappa_1 = \kappa e^{i\Delta\omega_L t}$. Introducing a Fourier transformation for both time $t$ and synthetic lattice index $n$, $u_n(t) = \frac{\omega_L}{2\pi} \int_{-\pi/\omega_L}^{\pi/\omega_L} d\tilde{k} \int d\tilde{\omega}\, u_{\tilde{k}}(\tilde{\omega}) e^{i(n\tilde{k}\omega_L - \tilde{\omega}t)}$, we find the relation

$$\tilde{\omega} = -2|\kappa| \sin(\tilde{k}\omega_L + \phi_0 + \Delta\omega_L t), \tag{5}$$

in which $\omega_L$ acts as the synthetic lattice constant and $|\kappa| = \frac{ek_0 A_0}{2\gamma m}$ is the coupling strength. For $\Delta\omega_L = 0$, Eq. (5) gives a dispersion relation, which is presented in Fig. 1e for $\phi_0 = \frac{\pi}{2}$. Note that the synthetic frequency $\tilde{\omega}$ has units of frequency, and the synthetic wave vector $\tilde{k}$ has units of time. As a result, the synthetic group velocity is

$$\tilde{v}_g(t) = \frac{\partial \tilde{\omega}}{\partial \tilde{k}} = -2|\kappa|\omega_L \cos(\tilde{k}\omega_L + \phi_0 + \Delta\omega_L t), \tag{6}$$

Correspondingly, the synthetic displacement in the PINEM spectrum (the accumulated energy transfer) is

$$\overline{\Delta\omega}(t) = \int_0^t \tilde{v}_g(t)dt = \frac{2|\kappa|\omega_L}{\Delta\omega_L}\left(\sin(\tilde{k}\omega_L + \phi_0) - \sin(\tilde{k}\omega_L + \phi_0 + \Delta\omega_L t)\right), \tag{7}$$

This yields Bloch oscillations in the PINEM lattice for ultrafast free electrons. The Bloch frequency corresponds to the detuning $\Omega_B = \Delta\omega_L = \omega_L - v_0 q$. Thus, the initial PINEM spectrum repeats itself after a cycle of the oscillation $T_B = 2\pi/\Omega_B$.

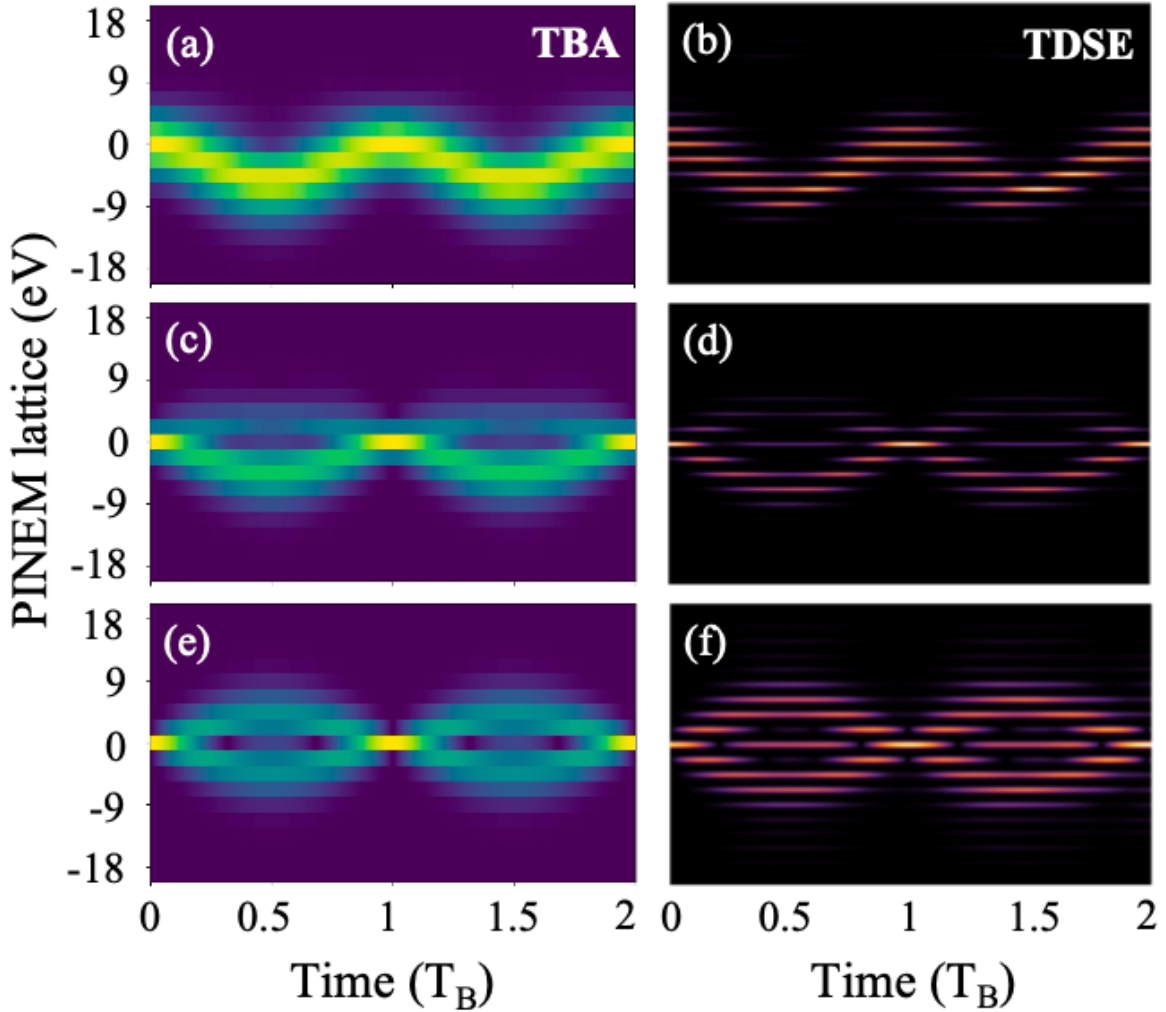

**Fig. 2: Demonstration of Bloch oscillation and breathing in the synthetic PINEM lattice.** The left column (a,c,e) is derived from the tight-binding-approximated model (TBA), showing the PINEM electron evolution along the energy axis. The electron exhibits (a) Bloch oscillation for the multiple sidebands within a Gaussian envelop and (c) Bloch breathing for single sideband input. The detuning ($\Delta\omega_L$) between the optical cycle and the grating period determines the Bloch period ($T_B$). Correspondingly, the right column (b, d, f) is derived from the time-dependent Schrödinger equation (TDSE). The PINEM pattern reveals Bloch oscillation in (b) and Bloch breathing in (d), both of which fit well with the TBA results. (c, d) both show the transition from Bloch oscillation to breathing. The frequency detuning $\Delta\omega_L = 1\,\mathrm{PHz}$, $T_B = 6.3\,\mathrm{fs}$, $\hbar\omega_L = 1.2\,\mathrm{eV}$, and $\sigma_E = 0.15\,\mathrm{eV}$.

**Bloch oscillation and breathing** – To numerically validate these analytical results, we compare the PINEM electron's evolution with the tight-binding-approximated (TBA) model (3) and the exact time-dependent Schrödinger equation (TDSE) (1). The TBA simulations are shown in Fig. 2a, 2c, 2e, and the corresponding TDSE results are shown in Fig. 2b, 2d, 2f. We can see that the transition from Bloch oscillation to Bloch breathing agrees well with the two approaches. The details of the analytical and numerical simulations are included in the SM file. Bloch breathing in strong-field physics has been previously reported[44].

The oscillation feature requires an initial electron with wide PINEM bandwidth, so that in the synthetic Bloch space, the momentum width ($\Delta\tilde{k}$, see Fig. 1e) is small enough to track the semiclassical trajectory, $\Delta\tilde{k} \ll \pi/\omega_L$. By contrast, Bloch breathing requires a PINEM electron with narrow or single sidebands corresponding to $\Delta\tilde{k} \approx \pi/\omega_L$. To control the width $\Delta\tilde{k}$, we can pre-select the PINEM electrons using a tip-based near-field interaction, which serves as a quantum state preparation. The tip-modulated interaction with quantum light can produce PINEM electrons with the required sideband distributions[4,39].

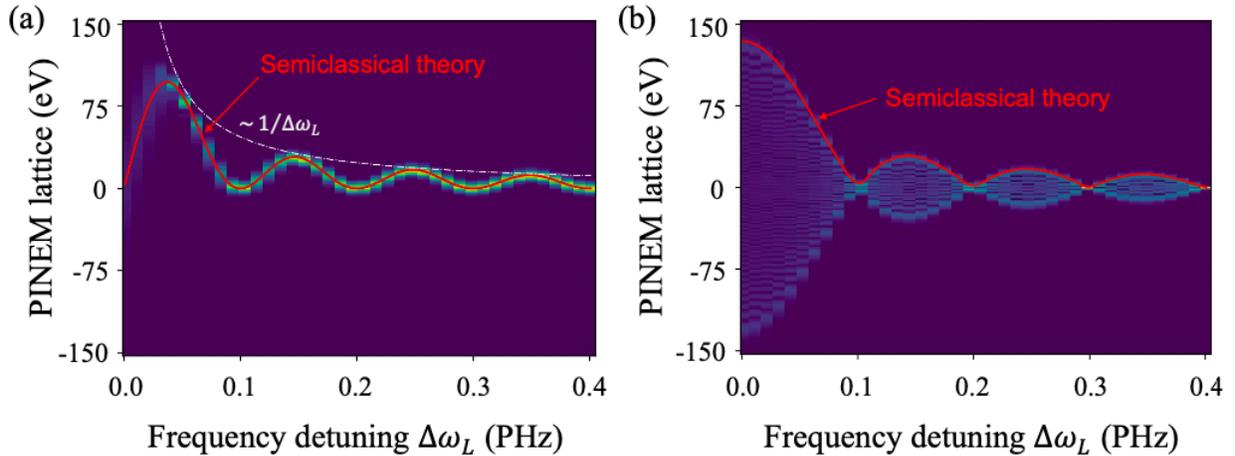

**Fig. 3: A proposal for verifying free-electron Bloch oscillations in experiment.** The semiclassical theory and TBA simulation match well for (a) the oscillating and (b) the breathing pattern. Both show an asymptotic behavior of $1/\Delta\omega_L$. The detuning $\Delta\omega_L$ can be readily altered by tuning the laser frequency. The interaction length L = 13 μm, the grating period Λ = 1.14 μm, the synchronized frequency $\omega_L^{(0)}$ = 2.36 PHz (corresponding to $\Delta\omega_L = 0$), and β = 0.7.

To verify the synthetic Bloch oscillation from EELS measurements, in experiment one can tune the frequency of the laser field to alter the detuning $\Delta\omega_L$. Fig. 3 depicts the oscillating and breathing spectrum as a function of $\Delta\omega_L$, in which the interaction length ($L = 13~\mu m$) and interaction field strength ($E_z = 10^8~V/m$) are kept fixed. Both the oscillation and asymptotic patterns as a function of the detuning can act as experimental signatures of the proposed synthetic Bloch oscillations.

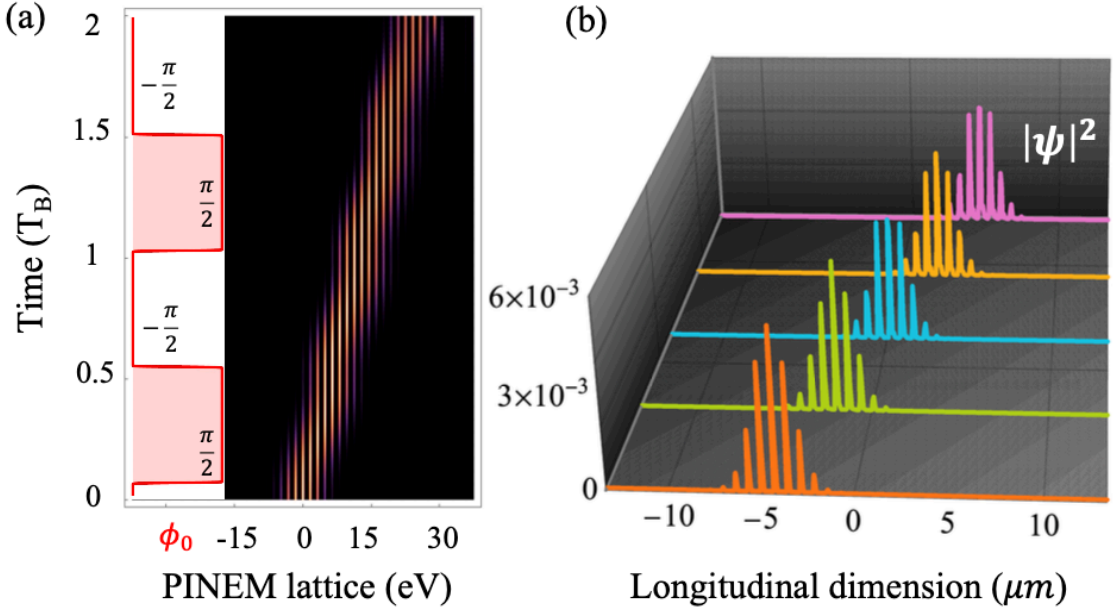

**Fig.4: Accelerating PINEM electrons based on synthetic Bloch oscillations.** (a) The evolution of PINEM electron throughout two oscillation periods $T_B$, with the phase of the modulation laser switching every half cycle, in which $T_B = 6.3$ fs in our simulation. Because of the periodically changing phase, the electron experiences an alternating effective electric force. As a result, the PINEM electron achieves a unidirectional energy transfer, which is comparable to the acceleration of a classical point-like electron. (b) The longitudinal (temporal) density distribution of the accelerated PINEM electron at $t = 0, T_B/2, T_B, 3T_B/2$, and $2T_B$, respectively.

**Accelerating PINEM electrons** – Now, we will control the flow of the synthetic Bloch oscillation (BO) for accelerating PINEM electrons. First, in the limit of synchronization $\omega_L/v_0 - q = 0$, i.e., $\Omega_B \rightarrow 0$, we find the averaged frequency shift of the PINEM electron,

$$\overline{\Delta\omega}\big|_{\Delta\omega_L \rightarrow 0} = -\frac{eE_0 L}{\hbar}\cos(\tilde{k}\omega_L + \phi_0),\tag{8}$$

in which the interaction length $L = v_0t$. Here, the expression $(-eE_0L)$ represents the classical work done on the electron. The phase $\tilde{k}\omega_L + \phi_0$ determines whether the PINEM electron accelerates or decelerates. This depends on the delay $\phi_0$ and, most strikingly, on the synthetic wave vector $\tilde{k}$ (see the band structure in the inset of Fig. 1e), which is determined by the PINEM state preparation. Only states in a specified range of synthetic wave vectors $\tilde{k}$ can be accelerated as given by Eq. 8, which is analogous to the diffraction management of light in photonic lattices[45,46] (see the discussion in the next section below).

On the other hand, for the phase-mismatched case ($\Delta\omega_L \neq 0$), there is no net acceleration for PINEM electrons in a long time-averaging ($t \gg T_B$). However, the oscillating PINEM distribution can achieve a peak spectral shift at half of the oscillation cycle ($T_B/2$), which is

$$\overline{\Delta\omega}\left(\frac{T_B}{2}\right) = \frac{2eE_0v_0T_B}{\hbar}\sin\left(\tilde{k}\omega_L + \phi_0\right),\qquad(10)$$

To alter this synthetic oscillation into a unidirectional shift, we can tune the oscillation direction after each half of the cycle. There are two ways to switch the phase. One simple trick is introducing a time-dependent delay $\phi_0(t)$ and sequentially tuning the delay from 0 to $\pi$ each half-period $t = T_B/2$. Figure 4a demonstrates the unidirectional transfer by this method. The alternative is to periodically change the sign of the detuning frequency $\pm\Delta\omega_L = \pm(\omega_L - v_0q)$, by merging two types of manufactured gratings with different periods in the propagation direction, whose detuning frequencies alternate between $+\Delta\omega_L$ and $-\Delta\omega_L$ at every distance of $v_0T_B/2$. As a result, the total energy transfer is $2N\overline{\Delta\omega}\left(\frac{T_B}{2}\right)$ within N Bloch cycles. This unidirectional spectral transfer has been discussed in dynamically modulated ring resonators[33].

We remark that accelerating PINEM electrons can also be understood by viewing the PINEM electrons as a train of attosecond pulses in the time domain that can be accelerated using beam techniques of accelerator physics [26,27]. As demonstrated in Fig. 4b, we present the longitudinal temporal profile of the accelerated PINEM electron in a train of attosecond pulses.

**Mimicking discrete linear optics.** We can take advantage of the close analogy between ultrafast electrons in the synthetic lattice and light propagation in a photonic lattice[45]. For instance, we can construct a synthetic diffraction coefficient $\tilde{D} = \frac{\partial^2\tilde{\omega}}{\partial\tilde{k}^2} = 2|\kappa|\omega_L^2\sin\left(\tilde{k}\omega_L + \phi_0\right)$ from the synthetic energy dispersion (5). As shown in Fig. 5a-5f, the discrete diffraction in the PINEM lattice can be normal or anomalous depending on the sign of the coefficient $\tilde{D}$, resulting in many noteworthy spectral patterns (broadening and focusing). For the case $\tilde{D} = 0$, we obtain the maximum group velocity $\tilde{v}_g = \pm 2|\kappa|\omega_L$, indicating that the PINEM electron can be accelerated or decelerated at the same rate as classical point particles in a laser-driven accelerator.

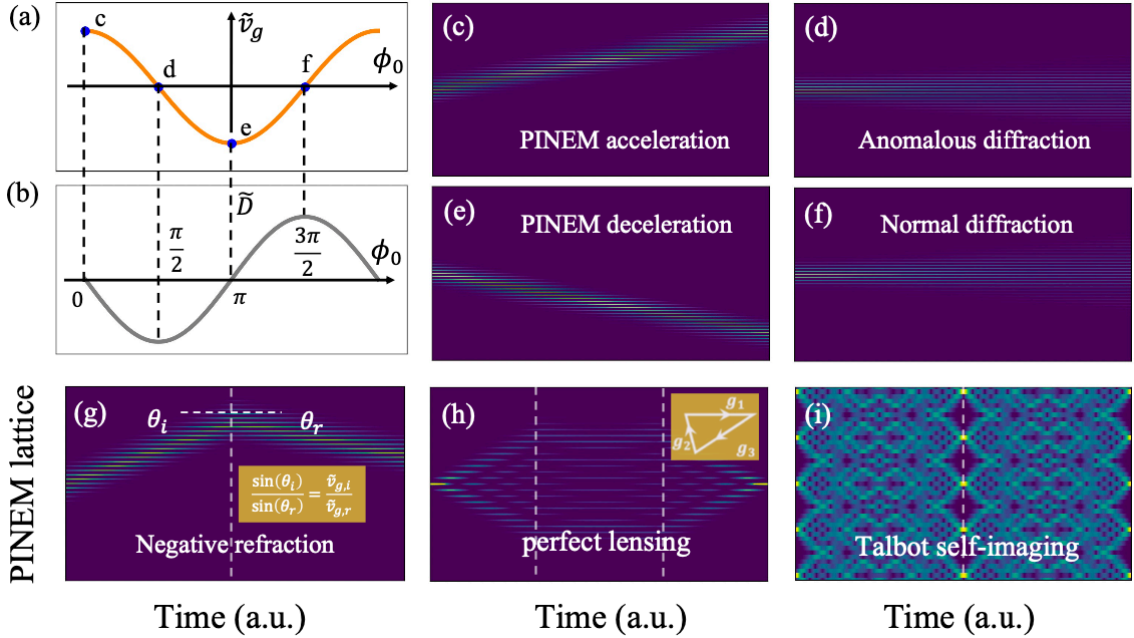

**Fig. 5: Mimicking diffraction management, negative refraction, perfect lensing, and Talbot self-imaging using PINEM electrons.** (a) The synthetic group velocity and (b) the diffraction coefficient as a function of the time delay ($\phi_0$). The spectral evolution of PINEM electrons undergoes (c) acceleration at $\phi_0 = 0$, (d) anomalous diffraction at $\phi_0 = \frac{\pi}{2}$, (e) deceleration at $\phi_0 = \pi$, and (f) normal diffraction at $\phi_0 = \frac{3\pi}{2}$, respectively. (g) The Snell's law for negative refraction is constructed using synthetic group velocities ($\sin\theta_i / \sin\theta_r = \tilde{v}_{g,1}/\tilde{v}_{g,2}$), and (h) the perfect lensing effect is constructed by phase control of the laser modulation in the condition ($\sum_i \mathbf{g}_i = 0$, as described in the SM file). (i) The spectral Talbot self-imaging is constructed using pre-selected initial PINEM electrons with the initial pattern (…1,0,-1,0,1,0,-1,0,…). The parameters are given in the SM file.

Also, many phenomena from linear or nonlinear optics, such as negative refraction, perfect lensing[47], and Talbot effect[48,49], can be easily engineered. Figure 5g-5i summarizes these discrete optical effects. For instance, in Fig. 5g, 5h, we use phase control of the laser modulation to obtain Snell's law for negative refraction and perfect lensing. The detailed analysis and parameters are given in the SM file. It is worth noting that the synthetic Talbot effect in Fig. 5i has a spectral character, in which the Talbot self-imaging occurs in energy space, in contrast to the temporal Talbot effect[18,50]. These optical phenomena, which are fascinating in and of themselves, can be utilized to manipulate and control ultrafast PINEM electrons.

**Further discussion** - Our proposal can be extended to construct high-dimensional synthetic PINEM lattices. For example, by using multicolor laser modulations with incommensurate frequencies[51], we can expand our simple tight-binding Hamiltonian (4) into a two-dimensional synthetic lattice $H = \sum_{\langle n,m \rangle} \kappa_{nm}(t) \left( e^{-i\phi_{nm}(t)} a_n^\dagger a_m + e^{i\phi_{nm}(t)} a_m^\dagger a_n \right)$, where the hopping coefficients $\kappa_{nm}$ (t) and phases $\phi_{nm}(t)$ between sidebands $n$ and $m$ can be controlled. Prospectively, state-of-the-art nanofabrication and laser techniques may enable the construction of artificial gauge fields and Floquet-engineering in synthetic dimensions[52], making it possible to explore the quantum Hall effect[53,54], topological phases[38,51,52], as well as quantum information processing[17–19] by PINEM electrons.

Note that we disregarded in this work the transverse dynamics of the electrons and the Coulomb interaction. However, the orbital angular momentum transverse to the propagation direction[55,56] might provide an additional extra dimension, which, combined with the longitudinal PINEM lattice, may allow for an alternate higher-dimensional synthetic space. Moreover, when the repulsive Coulomb forces between PINEM electrons are included, they induce interactions on the synthetic lattice Hamiltonian[32]. This interacting Hamiltonian may serve as an interesting starting point to explore many-body physics and nonlinear effects in ultrafast electrons.

In conclusion, we proposed a synthetic dimension using ultrafast PINEM electrons and demonstrated synthetic Bloch oscillations, PINEM acceleration, and diffraction management. These effects provide a proof-of-concept of PINEM synthetic dimensions as a novel simulation platform based on free electrons. The interplay between discrete linear optics and PINEM electrons opens up new opportunities for coherently manipulating free-electron beam dynamics in near-future experiments, such as UTEM, DLA, and QFEL systems.

## Methods

Methods, including modeling, calculation, and supplementary data, are available in the Supplementary information at [XXX].

## Acknowledgment

We thanks the discussions with Qingqing Cheng, Luqi Yuan, Eran Lustig, Maor Eldar. We thank the Israel Science Foundation for financial support (Grant No. 1803/18).

The authors declare no competing financial interests.

Correspondence and requests for materials should be addressed to Y.P (yiming.pan@campus.technion.ac.il) and B. Z (binzhang@mail.tau.ac.il).

# Supplementary Information

# Synthetic dimensions using ultrafast free electrons

Yiming Pan[1], Bin Zhang[2], Daniel Podolsky[1]

1. Department of Physics, Technion, Haifa 3200003, Israel

2. Department of Electrical Engineering Physical Electronics, Tel Aviv University, Ramat Aviv 6997801, Israel

This pdf file includes six sections and four supplementary figures.

## 1. Derivation of the TBA Hamiltonian from the TDSE (Eq. 4 in the main text).

Starting from the relativistically modified Schrodinger equation of a free electron in the presence of the electromagnetic field on a grating

$$i\hbar \frac{\partial}{\partial t}|\psi(t)\rangle = (\hat{H}_0 + \hat{H}_I)|\psi(t)\rangle \tag{1}$$

where the free-electron Hamiltonian is: $\hat{H}_0 = E_0 + v_0(\hat{p} - p_0) + \frac{(\hat{p}-p_0)^2}{2\gamma^3 m}$, with $E_0 = \gamma m c^2$, $p_0 = \gamma m v_0$, and $v_0 = \beta c$. The spatial wavefunction is defined as the projection of state on a spatial basis $\psi(z,t) = \langle z|\psi(t)\rangle$, and $\hat{p}$ is the momentum operator. The Lorentz factor $\gamma = 1/\sqrt{1-\beta^2}$. The interaction Hamiltonian between the electron and longitudinal near-field on the grating is given by: $\hat{H}_I = -\frac{e}{2\gamma m}[\hat{p} \cdot \hat{A}(\hat{z},t) + \hat{A}(\hat{z},t) \cdot p] = -\frac{e}{\gamma m}\left(\hat{A}(\hat{z},t) \cdot \hat{p} - \frac{i\hbar}{2}\frac{\partial}{\partial z}A(z,t)\right)$. The full 3D expression of laser-induced optical near-field for a DLA device can be found in Ref.[1] We notice that the Floquet harmonics on a grating are usually standing waves instead of traveling waves. Nevertheless, for the resonantly coupling with free electrons, only the forward-propagating components of Floquet modes can match the phase condition. So we consider the magnetic vector potential can be written as: $\hat{A}(\hat{z},t) = -\frac{E_0}{\omega_L}\sin(\omega_L t - q\hat{z} + \phi_0)$, with the electric field strength $E_0$, the laser frequency $\omega_L$, the gained wave vector $q$, and the phase delay $\phi_0$. Due to the spatial and temporal periodicities of the Hamiltonian, the free-electron wavefunction can be expanded on the Floquet-Bloch basis:

$$|\psi(t)\rangle = \sum_n a_n(z,t) e^{-i\omega_n t}|k_n\rangle \tag{2}$$

The basis $\{|k_n\rangle\}$ satisfies $\hat{p}|k_n\rangle = \hbar k_n|k_n\rangle$ and $\langle z|k_n\rangle = e^{ik_n z}$. with $\omega_n = \omega_0 + n\omega_L$ and $k_n = k_0 + nq$. $\omega_0$ and $k_0$ is the angular frequency and the wavevector of the input electron, satisfying $\hbar\omega_0 = E_0$ and $\hbar k_0 = p_0$. The index $n$ is an integer number. Substituting the electron wavefunction Eq. (2) into the Schrodinger equation Eq. (1), we get the LHS:

$$ i\hbar \frac{\partial}{\partial t}|\psi(t)\rangle = i\hbar \sum_n e^{-i\omega_n t}\left(-i\omega_n + \frac{\partial}{\partial t}\right)a_n(z,t)\,|k_n\rangle $$

$$ = \sum_n e^{-i\omega_n t}\left(\hbar\omega_n + i\hbar \frac{\partial}{\partial t}\right)a_n(z,t)\,|k_n\rangle \tag{3} $$

And the RHS contains two terms, the kinetic energy of the free electron

$$ \hat{H}_0|\psi(t)\rangle = \sum_n\left[E_0 + v_0(\hbar k_n - p_0) + \frac{(\hbar k_n - p_0)^2}{2\gamma^3 m} - i\hbar v_0 \frac{\partial}{\partial z}\right]a_n(z,t)e^{-i\omega_n t}|k_n\rangle $$

$$ = \sum_n\left[-i\hbar v_0 \frac{\partial}{\partial z}a_n(z,t) + \left[E_0 + v_0 n\hbar q + \frac{(n\hbar q)^2}{2\gamma^3 m}\right]a_n(z,t)\right]e^{-i\omega_n t}|k_n\rangle \tag{4} $$

The additional derivative term of t in Eq. 3 and of z in Eq. 4 can be rewritten in the co-moving frame $z' = z - v_0 t$

$$ \left(i\hbar \frac{\partial}{\partial t} + i\hbar v_0 \frac{\partial}{\partial z}\right)a_n(z,t) = \left(i\hbar \frac{\partial}{\partial t} + i\hbar v_0 \frac{\partial}{\partial z}\right)a_n(z - v_0 t, t) $$

$$ = i\hbar\left[\left(\frac{\partial}{\partial t} + \frac{\partial z'}{\partial t}\frac{\partial}{\partial z'}\right) + v_0 \frac{\partial z'}{\partial z}\frac{\partial}{\partial z'}\right]a_n(z - v_0 t, t) $$

$$ = i\hbar \frac{\partial}{\partial t}a_n(z', t) $$

Which means the amplitude can be written as the function of time: $a_n(z,t) \equiv a_n(t)$. Since the interaction term $\hat{H}_I$ contains both the coordinate $\hat{z}$ and momentum operator $\hat{p}$, we first insert the identity operator $\hat{I} = \int dz\,|z\rangle\langle z|$

$$\hat{H}_I |\psi(t)\rangle = -\frac{e}{\gamma m}\left(\hat{A}(\hat{z},t)\cdot\hat{p} - \frac{i\hbar}{2}\frac{\partial}{\partial z}A(z,t)\right)\sum_n a_n(z,t)\,e^{-i\omega_n t}|k_n\rangle$$

$$= -\hbar\frac{eA_0}{\gamma m}\sum_n\left[\left(k_n - i\frac{\partial}{\partial z}\right)a_n(z,t)\right]e^{-i\omega_n t}\int dz\,\sin(\omega_L t - q\hat{z} + \phi_0)\,|z\rangle\langle z|k_n\rangle$$

$$-i\hbar\frac{eA_0}{2\gamma m}q\cos(\omega_L t - q\hat{z} + \phi_0)\sum_n a_n(z,t)\,e^{-i\omega_n t}|k_n\rangle$$

$$= -\hbar\frac{eA_0}{\gamma m}\sum_n\left[\left(k_n - i\frac{\partial}{\partial z}\right)a_n(z,t)\right]e^{-i\omega_n t}\int dz\,e^{ik_n z}\sin(\omega_L t - qz + \phi_0)\,|z\rangle$$

$$-i\hbar\frac{eA_0}{2\gamma m}q\sum_n a_n(z,t)e^{-i\omega_n t}\int dz\,e^{ik_n z}\cos(\omega_L t - qz + \phi_0)\,|z\rangle$$

The sine and cosine functions can be written as

$$\sin(\omega_L t - qz + \phi_0) = -i\left[e^{i(\omega_L t - qz + \phi_0)} - e^{-i(\omega_L t - qz + \phi_0)}\right]$$
$$\cos(\omega_L t - q\hat{z} + \phi_0) = e^{i(\omega_L t - qz + \phi_0)} + e^{-i(\omega_L t - qz + \phi_0)}$$

Thus we get

$$\hat{H}_I|\psi(t)\rangle = i\hbar\frac{eA_0}{\gamma m}\sum_n\left[\left(k_n - i\frac{\partial}{\partial z}\right)a_n(z,t)\right]e^{-i\omega_n t}\int dz\left(e^{i(\omega_L t - k_{n-1}z + \phi_0)} - e^{-i(\omega_L t - k_{n+1}z + \phi_0)}\right)|z\rangle$$

$$-i\hbar\frac{eA_0}{2\gamma m}q\sum_n a_n(t)e^{-i\omega_n t}\int dz\,e^{i(\omega_L t - k_{n-1}z + \phi_0)} + e^{-i(\omega_L t - k_{n+1}z + \phi_0)}|z\rangle$$

$$= i\hbar\frac{eA_0}{\gamma m}\sum_n\left[\left(k_n - \frac{q}{2} - i\frac{\partial}{\partial z}\right)a_n(z,t)\right]\int dz\,e^{-i(\omega_{n-1}t - k_{n-1}z - \phi_0)}|z\rangle$$

$$-i\hbar\frac{eA_0}{\gamma m}\sum_n\left[\left(k_n + \frac{q}{2} - i\frac{\partial}{\partial z}\right)a_n(z,t)\right]\int dz\,e^{-i(\omega_{n+1}t - k_{n+1}z + \phi_0)}|z\rangle$$

(5)

Inserting the indentity operator $\hat{I} = \frac{1}{2\pi}\int dk\,|k\rangle\langle k|$ into the integration part

$$\int dz\, \hat{I} e^{-i(\omega_{n-1}t - k_{n-1}z - \phi_0)} |z\rangle = \frac{1}{2\pi} \int dz \int dk\, e^{-ikz} \left( e^{-i(\omega_{n-1}t - k_{n-1}z + \phi_0)} \right) |k\rangle$$

$$= \frac{1}{2\pi} \int dk \left( \int dz\, e^{i(k_{n-1}-k)z} \right) e^{-i(\omega_{n-1}t - \phi_0)} |k\rangle$$

$$= \int dk \left[ e^{-i(\omega_{n-1}t - \phi_0)} \delta(k_{n-1} - k) \right] |k\rangle$$

$$= e^{-i(\omega_{n-1}t - \phi_0)} |k_{n-1}\rangle$$

The same process will give

$$\int dz\, e^{-i(\omega_{n+1}t - k_{n+1}z + \phi_0)} |z\rangle = e^{-i(\omega_{n+1}t + \phi_0)} |k_{n+1}\rangle$$

Thus Eq. 5 can be written as

$$i\hbar \frac{eA_0}{\gamma m} \sum_n \left[ \left( k_n - \frac{q}{2} - i\frac{\partial}{\partial z} \right) a_n(z,t) \right] e^{-i(\omega_{n-1}t - \phi_0)} |k_{n-1}\rangle$$

$$- i\hbar \frac{eA_0}{\gamma m} \sum_n \left[ \left( k_n + \frac{q}{2} - i\frac{\partial}{\partial z} \right) a_n(z,t) \right] e^{-i(\omega_{n+1}t - \phi_0)} |k_{n+1}\rangle$$

$$= i\hbar \frac{eA_0}{\gamma m} \sum_n e^{-i\omega_n t} \left[ a_{n+1}(t) \left( k_{n+1} - \frac{q}{2} \right) e^{i\phi_0} - a_{n-1}(t) \left( k_{n-1} + \frac{q}{2} \right) e^{-i\phi_0} \right] |k_n\rangle$$

$$+ \hbar \frac{eA_0}{\gamma m} \sum_n e^{-i\omega_n t} \left[ e^{i\phi_0} \frac{\partial}{\partial z} a_{n+1}(z,t) - e^{-i\phi_0} \frac{\partial}{\partial z} a_{n-1}(z,t) \right] |k_n\rangle$$

Since we assume the amplitude $a_n(z,t)$ is a slow-varing function in both temporal and spatial domain, it is reasonable to drop the last term propotional second-order derivation ($\left[ e^{i\phi_0} \frac{\partial}{\partial z} a_{n+1}(z,t) - e^{-i\phi_0} \frac{\partial}{\partial z} a_{n-1}(z,t) \right]$). According to the relation $\omega_n = \omega_0 + n\omega_L$ and $k_n = k_0 + nq$ with $\hbar\omega_0 = E_0$ and $\hbar k_0 = p_0$, denote the comoving frame $t' = t - z/v_0$ as $t$, we get the coupled-mode equation

$$\left[ i\hbar \frac{\partial}{\partial t} + n\hbar \left( \omega_L - v_0 q - \frac{n\hbar q^2}{2\gamma^3 m} \right) \right] a_n(t)$$

$$= \hbar \frac{eA_0}{2\gamma m} \left[ \left( k_{n+1} - \frac{q}{2} \right) a_{n+1}(t) e^{i(\phi_0 + \pi/2)} + \left( k_{n-1} + \frac{q}{2} \right) a_{n-1}(t) e^{-i(\phi_0 + \pi/2)} \right]$$

(6)

Thus, we define the detuning frequency as

$$\Delta\omega_L = \omega_L - v_0 q - \frac{n\hbar q^2}{2\gamma^3 m} \simeq \omega_L - v_0 q \tag{7}$$

It is reasonable to treat $k_{n+1} - \frac{q}{2} \approx k_{n-1} + \frac{q}{2} \approx k_0$ since $q \ll k_0$ and define the coupling parameter between the neighboring mode as $\kappa = \frac{ek_0 A_0}{2\gamma m} e^{i(\phi_0 + \pi/2)}$. Hence, Eq. (6) can be written as

$$i\frac{\partial}{\partial t} a_n(t) = -n\Delta\omega_L a_n(t) + \kappa a_{n+1}(t) + \kappa^* a_{n-1}(t) \tag{8}$$

Thus, it is reasonable to define the slow varying state $|\psi_s(t)\rangle = \sum_n a_n(t) |n\rangle$, where $|n\rangle$ is the Wannier state in the synthetic space. We define $|\tilde{k}_\omega\rangle$ as the basis in the dual space of the synthetic space, which satisfies $\langle \tilde{k}_\omega | n \rangle = e^{ink_\omega \omega_L}$. It should be mentioned here that we have already worked on the new Hilbert space spanned by the basis $|n\rangle$, except for the same coefficient $a_n(t)$ as in Eq. (2). The tight-binding Hamiltonian for these synthetic ladders can be defined according to the coupled-mode equation

$$\begin{aligned}
\hat{H} &= -\Delta\omega_L \sum_n \mathrm{n} \, |\mathrm{n}\rangle\langle\mathrm{n}| + \sum_n \kappa \, |\mathrm{n}\rangle\langle\mathrm{n}+1| + h.c. \\
&= -\sum_n (n\Delta\omega_L)\hat{a}_n^\dagger \hat{a}_n + \sum_n \kappa \hat{a}_n^\dagger \hat{a}_{n+1} + h.c.
\end{aligned} \tag{9}$$

which satisfies the Schrodinger equation $i\frac{\partial}{\partial t}|\psi_s(t)\rangle = \hat{H}|\psi_s(t)\rangle$. Note that different from the photonic systems, here $a_n^\dagger, a_n$ are creation and annihilation operators, respectively, for fermions (i.e., electrons), with the anticommutation relation: $\{a_n^\dagger, a_m\} = \delta_{nm}$.

It's natural to define the Bloch representation $\{|\tilde{k}_\omega\rangle\}$ based on the Wannier basis $\{|n\rangle\}$

$$|\tilde{k}_\omega\rangle = \sum_n |n\rangle \langle n|\tilde{k}_\omega\rangle = \sqrt{\frac{\omega_L}{2\pi}} \sum_n e^{-in\tilde{k}_\omega \omega_L} |n\rangle$$

which satisfies Bloch condition $\langle n+1|\tilde{k}_\omega\rangle = e^{-i\tilde{k}_\omega \omega_L}\langle n|\tilde{k}_\omega\rangle$ with quasi-momentum $|\tilde{k}_\omega\rangle$ confined to the Brillouin zone $-\pi/\omega_L \leq \tilde{k}_\omega \leq \pi/\omega_L$. Then we could derive the tight-binding Hamiltonian in Bloch basis according to Eq. 9

$$\langle \tilde{k}'_\omega|\hat{H}|\tilde{k}_\omega\rangle = \langle \tilde{k}'_\omega| -\Delta\omega_L \sum_n n |n\rangle\langle n| + \sum_n \kappa |n\rangle\langle n+1| + h.c. |\tilde{k}_\omega\rangle$$

$$= -\Delta\omega_L \frac{\omega_L}{2\pi}\sum_n n\, e^{in(\tilde{k}'_\omega - \tilde{k}_\omega)\omega_L} + e^{-i\tilde{k}_\omega\omega_L}\frac{\omega_L}{2\pi}\sum_n \kappa\, e^{in(\tilde{k}'_\omega - \tilde{k}_\omega)\omega_L} + h.c.$$

$$= \delta\big(\tilde{k}'_\omega - \tilde{k}_\omega\big)\left[-i\frac{\Delta\omega_L}{\omega_L}\frac{d}{d\tilde{k}_\omega} + 2|\kappa|\cos\big(\tilde{k}_\omega\omega_L - \phi_0 - \pi/2\big)\right]$$

where $\delta\big(\tilde{k}'_\omega - \tilde{k}_\omega\big) = \frac{\omega_L}{2\pi}\sum_n e^{in(\tilde{k}'_\omega - \tilde{k}_\omega)\omega_L}$. Thus it's diagonal in the Bloch basis

$$H\big(\tilde{k}_\omega\big) = -i\frac{\Delta\omega_L}{\omega_L}\frac{d}{d\tilde{k}_\omega} + 2|\kappa|\sin\big(\tilde{k}_\omega\omega_L - \phi_0\big)$$

## 2. Wannier-Stark ladders using PINEM electrons

For a rigorous description of the coupling mode equation Eq. (8), we introduce the Wannier-Stark ladder representation. Expressing the state $|\psi_s(t)\rangle$ in the basis $|w_k\rangle$ with the time-independent coefficient $c_k$

$$|\psi_s(t)\rangle = \sum_n a_n(t)\,|n\rangle = \sum_k c_k\,e^{-i\Omega_k t}|w_k\rangle$$

where $\Omega_k$ is the eigenvalue corresponding to the eigenstate $|w_k\rangle$. Thus,

$$a_n(t) = \sum_k c_k\,e^{-i\Omega_k t} v_n^{(k)} \tag{10}$$

where the function $v_n^{(k)} = \langle k_n | w_k \rangle$ is the projection of the new basis $|w_k\rangle$ to the Bloch basis $|k_n\rangle$. With $\kappa = \dfrac{ek_0 A_0}{2\gamma m} e^{i(\phi_0 + \pi/2)}$, Eq. (8) can be written as

$$\Omega_k v_n^{(k)} = -n\Delta\omega_L v_n^{(k)} + |\kappa|e^{i(\phi_0+\pi/2)} v_{n+1}^{(k)} + |\kappa|e^{-i(\phi_0+\pi/2)} v_{n-1}^{(k)}$$

Redefine the function $\tilde{v}_n^{(k)} = v_n^{(k)} e^{in(\phi_0+\pi/2)}$, we can get a compact form

$$(\Omega_k + n\Delta\omega_L)\tilde{v}_n^{(k)} = |\kappa|\big[\tilde{v}_{n+1}^{(k)} + \tilde{v}_{n-1}^{(k)}\big] \tag{11}$$

Taking advantage of the recurrence relation of the Bessel functions

$$\frac{2n}{x} J_n(x) = J_{n+1}(x) + J_{n-1}(x) \tag{12}$$

Thus, the solution of Eq. (11) can be naturally derived by setting the eigenvalue $\Omega_k = -k\Delta\omega_L$, we can get the eigenfunction of Eq. (11) is

$$\tilde{v}_n^{(k)} = J_{n-k}\left(\frac{2|\kappa|}{\Delta\omega_L}\right)$$

$$a_n(t) = \sum_k c_k\,J_{n-k}\left(\frac{2|\kappa|}{\Delta\omega_L}\right) e^{ik\Delta\omega_L t}\,e^{-in(\phi_0+\pi/2)}$$

Therefore, the energy levels are equally spaced with energy difference $\Delta\omega_L$ between them. According to the orthonormality condition: $\sum_n J_{n-k}(x) J_{n-k'}(x) = \delta_{kk'}$,

$$\sum_n a_n(t)\left(v_n^{(k)}\right)^* = \sum_n a_n(t) J_{n-k}\left(\frac{2|\kappa|}{\Delta\omega_L}\right) e^{in(\phi_0 + \pi/2)}$$

$$= \sum_n J_{n-k}\left(\frac{2|\kappa|}{\Delta\omega_L}\right) \sum_{k'} c_{k'} J_{n-k'}\left(\frac{2|\kappa|}{\Delta\omega_L}\right) e^{-i\Omega_{k'}t}$$

$$= \sum_k c_{k'} e^{-i\Omega_{k'}t} \sum_n J_{n-k'}\left(\frac{2|\kappa|}{\Delta\omega_L}\right) J_{n-k}\left(\frac{2|\kappa|}{\Delta\omega_L}\right)$$

$$= c_k e^{-i\Omega_k t}$$

Thus, we can define the Wannier-Stark wavefunction

$$c_m = \sum_n a_n(0) J_{n-m}\left(\frac{2|\kappa|}{\Delta\omega_L}\right) e^{in(\phi_0 + \pi/2)} \tag{13}$$

where $c_m$ acts as the eigenstate with the eigenvalue $\Omega_m = -m\Delta\omega_L$, the energy spectrum is known as the Wannier-Stark ladder, characterized by the spectral localization in PINEM synthetic dimension. The Wannier-Stark wavefunction satisfies

$$\sum_m c_m^* c_m = \sum_{n,n'} a_n^*(0) a_{n'}(0) \sum_m J_{n-m}\left(\frac{2|\kappa|}{\Delta\omega_L}\right) J_{n'-m}\left(\frac{2|\kappa|}{\Delta\omega_L}\right)$$

$$= \sum_{n,n'} a_n^*(0) a_{n'}(0)\, \delta_{nn'} = \sum_n |a_n(0)|^2 = 1$$

Thus, the time evolution of the localized wavefunction $a_n(t)$ can be described through the Wannier-Stark state $c_m$,

$$a_n(t) = \sum_k \sum_{n'} a_{n'}(0) J_{n'-k}\left(\frac{2|\kappa|}{\Delta\omega_L}\right) J_{n-k}\left(\frac{2|\kappa|}{\Delta\omega_L}\right) e^{ik\Delta\omega_L t} e^{i(n-n')(\phi_0 + \pi/2)} \tag{14}$$

With the help of the identity formula: $\sum_n J_n(z) J_{n+p}(z)\, e^{in\alpha} = J_p\left[2z\sin\left(\frac{\alpha}{2}\right)\right] \exp[ip(\pi - \alpha)/2]$, and $e^{ig\sin\alpha} = \sum_n J_n(g)\, e^{in\alpha}$, the expression of $a_n(t)$ can be simplified as

$$
\begin{aligned}
a_n(t) &= \sum_{n'} a_{n'}(0) \sum_k J_{n'-k}\left(\frac{2|\kappa|}{\Delta\omega_L}\right) J_{n-k}\left(\frac{2|\kappa|}{\Delta\omega_L}\right) e^{ik\Delta\omega_L t} \, e^{i(n-n')(\phi_0+\pi/2)} \\
&= \sum_{n'} a_{n'}(0) \sum_{k'} J_{k'}\left(\frac{2|\kappa|}{\Delta\omega_L}\right) J_{k'+n-n'}\left(\frac{2|\kappa|}{\Delta\omega_L}\right) e^{-i(k'-n)\Delta\omega_L t} \, e^{i(n-n')(\phi_0+\pi/2)} \\
&= \sum_{n'} a_{n'}(0) \, J_{n'-n}\left[\frac{4|\kappa|}{\Delta\omega_L}\sin\left(\frac{\Delta\omega_L t}{2}\right)\right] \exp\left[-i(n-n')\left(\phi_0+\frac{\Delta\omega_L t}{2}\right)+in\Delta\omega_L t\right]
\end{aligned}
$$

Since the state $|\psi_s(t)\rangle$ could also be expressed on Bloch basis

$$
|\psi_s(t)\rangle = \sum_n a_n(t)\,|n\rangle = \int_{-\pi/\omega_L}^{\pi/\omega_L} d\tilde{k}_\omega \, a_{\tilde{k}_\omega}(t)\big|\tilde{k}_\omega\big\rangle
$$

thus we could define the Fourier transformation for $a_n(t)$

$$
\begin{aligned}
a_{\tilde{k}_\omega}(t) &= \sum_n a_n(t)\,\langle\tilde{k}_\omega|n\rangle = \sqrt{\frac{\omega_L}{2\pi}}\sum_n a_n(t)\,e^{in\tilde{k}_\omega\omega_L} \\
&= \sqrt{\frac{\omega_L}{2\pi}}\sum_{n',n} a_{n'}(0) J_{n'-n}\left[\frac{4|\kappa|}{\Delta\omega_L}\sin\left(\frac{\Delta\omega_L t}{2}\right)\right]\exp\left[i(n'-n)\left(\phi_0+\frac{\Delta\omega_L t}{2}\right)+in\Delta\omega_L t\right]e^{in\tilde{k}_\omega\omega_L} \\
&= \sqrt{\frac{\omega_L}{2\pi}}\sum_{n',n} a_{n'}(0) J_{n'-n}\left[\frac{4|\kappa|}{\Delta\omega_L}\sin\left(\frac{\Delta\omega_L t}{2}\right)\right]\exp\left[i(n-n')\left(\tilde{k}_\omega\omega_L+\frac{\Delta\omega_L t}{2}-\phi_0\right)\right]e^{in'(\tilde{k}_\omega\omega_L+\Delta\omega_L t)} \\
&= \sqrt{\frac{\omega_L}{2\pi}}\exp\left[i\frac{4|\kappa|}{\Delta\omega_L}\sin\left(\frac{\Delta\omega_L t}{2}\right)\sin\left(\tilde{k}_\omega\omega_L+\frac{\Delta\omega_L t}{2}-\phi_0\right)\right]\sum_{n'} a_{n'}(0)\,e^{in'(\tilde{k}_\omega\omega_L+\Delta\omega_L t)}
\end{aligned}
\tag{15}
$$

The averaged trajectory of the electron in the PINEM synthetic dimension is

$$
\langle x\rangle = \omega_L\langle n\rangle = \omega_L\sum_n n\,|a_n|^2
$$

And the width of the wavefunction distribution is defined by the variance

$$
\begin{aligned}
\Delta x = \langle x^2\rangle - \langle x\rangle^2 &= \omega_L^2(\langle n^2\rangle - \langle n\rangle^2) \\
&= \omega_L^2\left[\sum_n n^2\,|a_n|^2 - \left(\sum_n n\,|a_n|^2\right)^2\right]
\end{aligned}
$$

## 3. Synthetic band analysis for PINEM electrons

We develop another method to solve Eq. (8) by introducing the unitary transformation $a_n(t) = e^{in\Delta\omega_L t}u_n(t)$, the LHS of Eq. (8) becomes $e^{in\Delta\omega_L t}\left[i\frac{\partial}{\partial t}u_n(t) - n\Delta\omega u_n(t)\right]$. And the corresponding RHS is

$$-n\Delta\omega_L u_n(t)e^{in\Delta\omega_L t} + \kappa e^{i(n+1)\Delta\omega_L t}u_{n+1}(t) + \kappa^* e^{i(n-1)\Delta\omega_L t}u_{n-1}(t)$$
$$= e^{in\Delta\omega_L t}\left[-n\Delta\omega_L u_n(t) + \kappa e^{i\Delta\omega_L t}u_{n+1}(t) + \kappa^* e^{-i\Delta\omega_L t}u_{n-1}(t)\right] \tag{16}$$

Thus, we get a simplified coupled mode equation

$$i\frac{\partial}{\partial t}u_n(t) = \kappa_1 u_{n+1}(t) + \kappa_1^* u_{n-1}(t) \tag{17}$$

where $\kappa_1 = \kappa e^{i\Delta\omega_L t} = \frac{ek_0A_0}{2\gamma m}e^{i\left(\phi_0 + \frac{\pi}{2} + \Delta\omega t\right)}$. We could derive the relation of the Fourier component $a_{\tilde{k}_\omega}(t)$ and $\tilde{u}_{\tilde{k}_\omega}(t)$ according to the relation $a_n(t) = e^{in\Delta\omega_L t}u_n(t)$

$$\begin{aligned}
\tilde{u}_{\tilde{k}_\omega}(t) &= \sqrt{\frac{\omega_L}{2\pi}}\sum_n u_n(t)\,e^{in\tilde{k}_\omega\omega_L} = \sqrt{\frac{\omega_L}{2\pi}}\sum_n a_n(t)\,e^{in(\tilde{k}_\omega\omega_L - \Delta\omega_L t)} \\
&= \frac{\omega_L}{2\pi}\sum_n \int_{-\pi/\omega_L}^{\pi/\omega_L} a_{\tilde{k}}(t)\,e^{-in\tilde{k}\omega_L}d\tilde{k}\,e^{in(\tilde{k}_\omega\omega_L - \Delta\omega_L t)} \\
&= \frac{\omega_L}{2\pi}\int_{-\pi/\omega_L}^{\pi/\omega_L} d\tilde{k}\,a_{\tilde{k}}(t)\sum_n e^{in\omega_L(\tilde{k}_\omega - \tilde{k})}\,e^{-in\Delta\omega_L t} \\
&= \int_{\frac{\pi}{\omega_L}}^{\frac{\pi}{\omega_L}} d\tilde{k}\,a_{\tilde{k}}(t)\,\delta\left(\omega_L(\tilde{k}_\omega - \tilde{k}) - \Delta\omega_L t\right) \\
&= a_{\tilde{k}_\omega - \Delta\omega_L t/\omega_L}(t)
\end{aligned}$$

This procedure is similar to the canonical substitution of a wave packet in the presence of an electromagnetic field, i.e., $\tilde{k}_\omega \to \tilde{k}_\omega - \Delta\omega t/\omega_L$. So that the term $\Delta\omega t/\omega_L$ mimics an effective magnetic vector potential $\tilde{A}(t) = \Delta\omega t/\omega_L$.

Introducing the Fourier transformation for time t, we get the expansion

$$u_n(t) = \sqrt{\frac{\omega_L}{2\pi}} \int_{-\frac{\pi}{\omega_L}}^{\frac{\pi}{\omega_L}} d\tilde{k}_\omega \int d\omega \, \tilde{u}_{\tilde{k}_\omega}(\omega) e^{-i\omega t} e^{-in\tilde{k}_\omega \omega_L} \tag{18}$$

Insert the above equation into Eq. (17), and we get the identity

$$\sqrt{\frac{\omega_L}{2\pi}} \int_{-\frac{\pi}{\omega_L}}^{\frac{\pi}{\omega_L}} d\tilde{k}_\omega \int d\omega \left[ \omega - \left( \kappa_1 e^{-i\tilde{k}_\omega \omega_L} + \kappa_1^* e^{i\tilde{k}_\omega \omega_L} \right) \right] \tilde{u}_{\tilde{k}_\omega}(\omega) e^{-i(n\tilde{k}_\omega \omega_L + \omega t)} = 0$$

which implies $\omega - \left( \kappa_1 e^{-i\tilde{k}_\omega \omega_L} + \kappa_1^* e^{i\tilde{k}_\omega \omega_L} \right) \equiv 0$. Thus, we get the dispersion relation in the time domain for the system

$$\omega = -2|\kappa_1| \sin\left( \tilde{k}_\omega \omega_L - \phi_0 - \Delta\omega t \right) \tag{19}$$

where $\omega_L$ is the synthetic lattice constant. Here we view that the coupling $\kappa_1$ adiabatically follows the evaluation time.

For any input mode distribution $u_n(0)$, we can get the distribution $\tilde{u}_{\tilde{k}_\omega}(t) = \sum_n u_n(t) \, e^{in\tilde{k}_\omega \omega_L}$ in the synthetic momentum space $\tilde{k}_\omega$, according to the Fourier transformation defined in Eq. (18). The evolution of $\tilde{u}_{k_\omega}(t)$ can be derived from Eq. (17):

$$i \frac{\partial}{\partial t} \tilde{u}_{\tilde{k}_\omega}(t) = \left( \kappa_1 \, e^{-i\tilde{k}_\omega \omega_L} + \kappa_1^* e^{i\tilde{k}_\omega \omega_L} \right) \tilde{u}_{\tilde{k}_\omega}(t)$$
$$= -2|\kappa| \sin\left( \tilde{k}_\omega \omega_L - \Delta\omega t - \phi_0 \right) \tilde{u}_{\tilde{k}_\omega}(t) \tag{20}$$

Integration over time on both sides, we get

$$\int_0^{t_f} dt \frac{\partial_t \tilde{u}_{k_\omega}(t)}{\tilde{u}_{k_\omega}(t)} = \ln\left[ \frac{\tilde{u}_{k_\omega}(t_f)}{\tilde{u}_{k_\omega}(0)} \right] = 2i|\kappa| \int_0^{t_f} dt \sin\left( \tilde{k}_\omega \omega_L - \Delta\omega t - \phi_0 \right)$$
$$= -\frac{2i|\kappa|}{\Delta\omega} \left[ \cos\left( \tilde{k}_\omega \omega_L - \Delta\omega t_f - \phi_0 \right) - \cos\left( \tilde{k}_\omega \omega_L - \phi_0 \right) \right]$$
$$= \frac{4i|\kappa|}{\Delta\omega} \sin\left( \frac{\Delta\omega t_f}{2} \right) \sin\left( \tilde{k}_\omega \omega_L - \phi_0 - \frac{\Delta\omega t_f}{2} \right)$$

Thus

$$\tilde{u}_{\tilde{k}_\omega}(t_f) = \tilde{u}_{\tilde{k}_\omega}(0) \exp\left[i\frac{4|\kappa|}{\Delta\omega}\sin\left(\frac{\Delta\omega t_f}{2}\right)\sin\left(\tilde{k}_\omega\omega_L - \phi_0 - \frac{\Delta\omega t_f}{2}\right)\right]$$

$$= \tilde{u}_{\tilde{k}_\omega}(0) \sum_n J_n\left[\frac{4|\kappa|}{\Delta\omega}\sin\left(\frac{\Delta\omega t_f}{2}\right)\right] e^{in\left(\tilde{k}_\omega\omega_L - \phi_0 - \frac{\Delta\omega t_f}{2}\right)} \tag{21}$$

And the mode distribution at arbitrary time $t$ is

$$u_n(t) = \sqrt{\frac{\omega_L}{2\pi}} \int_{-\frac{\pi}{\omega_L}}^{\frac{\pi}{\omega_L}} \tilde{u}_{\tilde{k}_\omega}(t)\, e^{-in\tilde{k}_\omega\omega_L} d\tilde{k}_\omega$$

$$= \sqrt{\frac{\omega_L}{2\pi}} \sum_m J_m\left[\frac{4|\kappa|}{\Delta\omega}\sin\left(\frac{\Delta\omega t_f}{2}\right)\right]$$

$$\times \int_{-\frac{\pi}{\omega_L}}^{\frac{\pi}{\omega_L}} \tilde{u}_{\tilde{k}_\omega}(0)\, e^{im\left(\tilde{k}_\omega\omega_L - \phi_0 - \frac{\Delta\omega t_f}{2}\right)} e^{-in\tilde{k}_\omega\omega_L} d\tilde{k}_\omega \tag{22}$$

Now we consider going back to the Wannier representation, according to $a_n(t) = e^{in\Delta\omega_L t} u_n(t)$ and the Fourier transformation

$$\tilde{u}_{\tilde{k}_\omega}(0) = \sqrt{\frac{\omega_L}{2\pi}} \sum_n u_n(0)\, e^{in\tilde{k}_\omega\omega_L} = \sqrt{\frac{\omega_L}{2\pi}} \sum_n a_n(0)\, e^{in\tilde{k}_\omega\omega_L}$$

According to the first equality of Eq. (21),

$$a_{\tilde{k}_\omega - \frac{\Delta\omega_L}{\omega_L}t}(t) = \exp\left[i\frac{4|\kappa|}{\Delta\omega}\sin\left(\frac{\Delta\omega t}{2}\right)\sin\left(\tilde{k}_\omega\omega_L - \phi_0 - \frac{\Delta\omega t}{2}\right)\right] \sum_n a_n(0)\, e^{in\tilde{k}_\omega\omega_L}$$

Then we can replicate the conclusion Eq. (15) derived by the Wannier-Stark ladders

$$a_{\tilde{k}_\omega}(t) = \exp\left[i\frac{4|\kappa|}{\Delta\omega}\sin\left(\frac{\Delta\omega t}{2}\right)\sin\left(\tilde{k}_\omega\omega_L - \phi_0 + \frac{\Delta\omega t}{2}\right)\right] \sum_n a_n(0)\, e^{in(\tilde{k}_\omega\omega_L + \Delta\omega t)} \tag{23}$$

## 4. Bloch breathing and oscillation in synthetic space

For the input state $a_n(0) = u_n(0) = \delta_{n0}$, we can obtain

$$\tilde{u}_{\tilde{k}_\omega}(0) = \sqrt{\frac{\omega_L}{2\pi}} \sum_n \delta_{n0}\, e^{in\tilde{k}_\omega \omega_L} = \sqrt{\frac{\omega_L}{2\pi}}$$

According to Eq. (22), the mode distribution at arbitrary time $t$ is

$$
\begin{aligned}
u_n(t) &= \frac{\omega_L}{2\pi} \sum_m J_m\left[\frac{4|\kappa|}{\Delta\omega}\sin\left(\frac{\Delta\omega t}{2}\right)\right] e^{-im\left(\phi_0 + \frac{\Delta\omega t}{2}\right)} \int_{-\pi}^{\pi} e^{-i(n-m)\tilde{k}_\omega \omega_L}\, d\tilde{k}_\omega \\
&= \sum_m J_m\left[\frac{4|\kappa|}{\Delta\omega}\sin\left(\frac{\Delta\omega t}{2}\right)\right] e^{-im\left(\phi_0 + \frac{\Delta\omega t}{2}\right)} \delta_{nm} \\
&= J_n\left[\frac{4|\kappa|}{\Delta\omega}\sin\left(\frac{\Delta\omega t}{2}\right)\right] e^{-in\left(\phi_0 + \frac{\Delta\omega t}{2}\right)}.
\end{aligned}
\tag{24}
$$

Thus $a_n(t) = e^{in\Delta\omega_L t} u_n(t) = J_n\left[\frac{4|\kappa|}{\Delta\omega}\sin\left(\frac{\Delta\omega t}{2}\right)\right] e^{-in\left(\phi_0 - \frac{\Delta\omega t}{2}\right)}$, consistent with the result shown in the main text Fig. 2e, 2f. We can recover the conventional PINEM spectrum in the limit $\Delta\omega \to 0$. Since the component inside the Bessel function follows

$$\lim_{\Delta\omega \to 0} \frac{4|\kappa|}{\Delta\omega}\sin\left(\frac{\Delta\omega t}{2}\right) = 2|\kappa|t.$$

Then the mode distribution after modulation for the single-mode input $a_n(0) = \delta_{n0}$ is

$$a_n(t) = J_n(2|\kappa|t)e^{-in\phi_0},$$

which is the conventional PINEM distribution.

According to the properties of the Bessel function, $J_{-n}(\alpha) = (-1)^n J_n(\alpha)$ and Eq. 12, we obtain that the trajectory of the PINEM distribution is

$$\langle x(t) \rangle = \omega_L \sum_{n=-\infty}^{+\infty} n \, |a_n|^2 = \omega_L \sum_{n=-\infty}^{+\infty} n \left| J_n \left[ \frac{4|\kappa|}{\Delta\omega} \sin\left(\frac{\Delta\omega t}{2}\right) \right] \right|^2$$

$$= \omega_L \sum_{n=1}^{+\infty} \left[ n \left| J_n \left[ \frac{4|\kappa|}{\Delta\omega} \sin\left(\frac{\Delta\omega t}{2}\right) \right] \right|^2 - n \left| J_{-n} \left[ \frac{4|\kappa|}{\Delta\omega} \sin\left(\frac{\Delta\omega t}{2}\right) \right] \right|^2 \right] \qquad (25)$$

$$+ \omega_L \times 0 \times \left| J_0 \left[ \frac{4|\kappa|}{\Delta\omega} \sin\left(\frac{\Delta\omega t}{2}\right) \right] \right|^2 = 0.$$

and the variance is

$$\langle x^2 \rangle = \omega_L^2 \sum_{n=-\infty}^{+\infty} n^2 \, |a_n|^2 = \omega_L^2 \sum_{n=-\infty}^{+\infty} \left[ n J_n \left[ \frac{4|\kappa|}{\Delta\omega} \sin\left(\frac{\Delta\omega t}{2}\right) \right] \right]^2$$

$$= \omega_L^2 \left[ \frac{2|\kappa|}{\Delta\omega} \sin\left(\frac{\Delta\omega t}{2}\right) \right]^2 \sum_{n=-\infty}^{+\infty} \left[ \left( J_{n+1} \left[ \frac{4|\kappa|}{\Delta\omega} \sin\left(\frac{\Delta\omega t}{2}\right) \right] + J_{n-1} \left[ \frac{4|\kappa|}{\Delta\omega} \sin\left(\frac{\Delta\omega t}{2}\right) \right] \right) \right]^2 \qquad (26)$$

$$= \frac{8|\kappa|^2 \omega_L^2}{\Delta\omega^2} \sin^2\left(\frac{\Delta\omega t}{2}\right) = \frac{4|\kappa|^2 \omega_L^2}{\Delta\omega^2} \left[ 1 - \cos(\Delta\omega t) \right].$$

So, we notice that the non-zero variance $\langle x^2 \rangle$ oscillates with a period of $T_B = \frac{2\pi}{\Delta\omega}$.

On the other hand, for the input PINEM state $a_n(0) = u_n(0) = \left( \frac{\omega_L^2}{2\pi\sigma_{en}^2} \right)^{1/4} \exp\left[ -\frac{n^2 \omega_L^2}{4\sigma_{en}^2} \right]$, under the assumption $\sigma_{en} \gg \omega_L$, the summation in Eq. (23) can be replaced by an integration,

$$\sum_n a_n(0) \, e^{in(\bar{k}_\omega \omega_L + \Delta\omega t)} = \left( \frac{\omega_L^2}{2\pi\sigma_{en}^2} \right)^{\frac{1}{4}} \sum_n \exp\left[ -\frac{(n\omega_L)^2}{4\sigma_{en}^2} \right] e^{in\omega_L\left(\bar{k}_\omega + \frac{\Delta\omega}{\omega_L} t\right)}$$

$$= \left( \frac{\omega_L^2}{2\pi\sigma_{en}^2} \right)^{\frac{1}{4}} \sum_n \Delta x \exp\left[ -\frac{(n\Delta x \omega_L)^2}{4\sigma_{en}^2} \right] e^{in\Delta x \omega_L\left(\bar{k}_\omega + \frac{\Delta\omega}{\omega_L} t\right)}$$

$$= \left( \frac{\omega_L^2}{2\pi\sigma_{en}^2} \right)^{\frac{1}{4}} \int_{-\infty}^{\infty} dx \exp\left[ -\frac{\omega_L^2 x^2}{4\sigma_{en}^2} \right] e^{i\omega_L x\left(\bar{k}_\omega + \frac{\Delta\omega}{\omega_L} t\right)}$$

$$= \left( \frac{8\pi\sigma_{en}^2}{\omega_L^2} \right)^{\frac{1}{4}} \exp\left[ -\sigma_{en}^2 \left( k + \frac{\Delta\omega}{\omega_L} t \right)^2 \right].$$

Thus, the distribution in the synthetic momentum space is

$$
\begin{aligned}
a_{\tilde{k}_\omega}(t) &= \exp\left[i\,\frac{4|\kappa|}{\Delta\omega}\sin\left(\frac{\Delta\omega t}{2}\right)\sin\left(\tilde{k}_\omega\omega_L - \phi_0 + \frac{\Delta\omega t}{2}\right)\right]\sum_n a_n(0)\,e^{in(\tilde{k}_\omega\omega_L + \Delta\omega t)} \\
&= \exp\left[i\,\frac{4|\kappa|}{\Delta\omega}\sin\left(\frac{\Delta\omega t}{2}\right)\sin\left(\tilde{k}_\omega\omega_L - \phi_0 + \frac{\Delta\omega t}{2}\right)\right] \\
&= \left(\frac{8\pi\sigma_{en}^2}{\omega_L^2}\right)^{\frac{1}{4}}\exp\left[i\,\frac{4|\kappa|}{\Delta\omega}\sin\left(\frac{\Delta\omega t}{2}\right)\sin\left(\tilde{k}_\omega\omega_L - \phi_0 + \frac{\Delta\omega t}{2}\right)\right] \\
&\quad\times \exp\left[-\sigma_{en}^2\left(\tilde{k}_\omega + \frac{\Delta\omega}{\omega_L}t\right)^2\right]
\end{aligned}
\tag{27}
$$

Since the wavepacket distribution is widely spread in the synthetic space, i.e., $\sigma_{en} \gg \omega_L$, it has a very localized synthetic wavevector in its synthetic Bloch space. This offers us a chance to expand the sine function $\sin\left(\tilde{k}_\omega\omega_L - \phi_0 + \frac{\Delta\omega t}{2}\right)$ around $\tilde{k}_\omega = -\frac{\Delta\omega}{\omega_L}t$ in the phase term of Eq. (27) to the first order,

$$
\sin\left(\tilde{k}_\omega\omega_L - \phi_0 + \frac{\Delta\omega t}{2}\right) \simeq -\sin\left(\phi_0 + \frac{\Delta\omega t}{2}\right) + \omega_L\cos\left(\phi_0 + \frac{\Delta\omega t}{2}\right)\left(\tilde{k}_\omega + \frac{\Delta\omega}{\omega_L}t\right)
$$

Thus, the phase term becomes

$$
\exp\left[i\,\frac{4|\kappa|}{\Delta\omega}\sin\left(\frac{\Delta\omega t}{2}\right)\sin\left(\tilde{k}_\omega\omega_L - \phi_0 + \frac{\Delta\omega t}{2}\right)\right] = \exp\left[i\,\frac{4|\kappa|}{\Delta\omega}\left[\Phi(t) + d(t)\left(\tilde{k}_\omega\omega_L + \Delta\omega t\right)\right]\right]
$$

where

$$
\Phi(t) = -\sin\left(\frac{\Delta\omega t}{2}\right)\sin\left(\phi_0 + \frac{\Delta\omega t}{2}\right) = \frac{1}{2}\left[\cos(\Delta\omega t + \phi_0) - \cos(\phi_0)\right]
$$

$$
d(t) = \sin\left(\frac{\Delta\omega t}{2}\right)\cos\left(\phi_0 + \frac{\Delta\omega t}{2}\right) = \frac{1}{2}\left[\sin(\Delta\omega t + \phi_0) + \sin(\phi_0)\right]
$$

Thus, we get

$$a_n(t) = \frac{\omega_L}{2\pi} \int_{-\frac{\pi}{\omega_L}}^{\frac{\pi}{\omega_L}} d\tilde{k}_\omega \, a_{\tilde{k}_\omega}(t) \, e^{in\tilde{k}_\omega \omega_L} = \frac{\omega_L}{2\pi} \left( \frac{8\pi\sigma_{en}^2}{\omega_L^2} \right)^{\frac{1}{4}} \exp\left[ i\frac{4|\kappa|}{\Delta\omega} \Phi(t) \right]$$

$$\times \int_{-\frac{\pi}{\omega_L}}^{\frac{\pi}{\omega_L}} d\tilde{k}_\omega \, e^{in\tilde{k}_\omega \omega_L} \exp\left[ i\frac{4|\kappa|}{\Delta\omega} \omega_L d(t) \left( \tilde{k}_\omega + \frac{\Delta\omega}{\omega_L} t \right) \right] \exp\left[ -\sigma_{en}^2 \left( \tilde{k}_\omega + \frac{\Delta\omega}{\omega_L} t \right)^2 \right] \quad (28)$$

$$= \left( \frac{\omega_L^2}{2\pi\sigma_{en}^2} \right)^{1/4} \exp\left[ -\frac{1}{4\sigma_{en}^2} \left( n\omega_L + \frac{4|\kappa|}{\Delta\omega} \omega_L d(t) \right)^2 - in\Delta\omega t + i\frac{4|\kappa|}{\Delta\omega} \Phi(t) \right]$$

This result is fully consistent with the TBA results, as shown in Fig. 2a, 2c, 2e in the main text, which is numerically calculated through Eq. (8), with the detuning $\Delta\omega = 1 \, fs^{-1}$ and the coupling constant $|\kappa| = 0.7$, corresponding to the laser field $E_0 \simeq 7 \times 10^7 \, V/m$.

Then the trajectory is

$$\langle x \rangle = \omega_L \sum_n n \, |a_n|^2 = \frac{\omega_L^2}{\sqrt{2\pi\sigma_{en}^2}} \sum_n n \exp\left[ -\frac{1}{2\sigma_{en}^2} \left( n\omega_L + \frac{4|\kappa|}{\Delta\omega} \omega_L d(t) \right)^2 \right]$$

$$= \frac{\omega_L^2}{\sqrt{2\pi\sigma_{en}^2}} \int dx \, x \exp\left[ -\frac{\omega_L^2}{2\sigma_{en}^2} \left( x + \frac{4|\kappa|}{\Delta\omega} d(t) \right)^2 \right] \quad (29)$$

$$= \frac{4|\kappa|}{\Delta\omega} \omega_L d(t),$$

and

$$\langle x^2 \rangle = \omega_L^2 \sum_n n^2 |a_n|^2 = \frac{\omega_L^3}{\sqrt{2\pi\sigma_{en}^2}} \sum_n n^2 \exp\left[ -\frac{1}{2\sigma_{en}^2}\left( n\omega_L + \frac{4|\kappa|}{\Delta\omega}\omega_L d(t) \right)^2 \right]$$

$$= \frac{\omega_L^3}{\sqrt{2\pi\sigma_{en}^2}} \int dx\, x^2 \exp\left[ -\frac{\omega_L^2}{2\sigma_{en}^2}\left( x + \frac{4|\kappa|}{\Delta\omega} d(t) \right)^2 \right]$$

$$= \frac{\omega_L^3}{\sqrt{2\pi\sigma_{en}^2}} \int dx \left[ x^2 + \frac{16|\kappa|^2}{\Delta\omega^2} d(t)^2 \right] \exp\left[ -\frac{\omega_L^2}{2\sigma_{en}^2} x^2 \right]$$

$$= \frac{\omega_L^3}{\sqrt{2\pi\sigma_{en}^2}} \left[ 2\sqrt{2\pi}\frac{\sigma_{en}^3}{\omega_L^3} + \frac{16|\kappa|^2}{\Delta\omega^2} d(t)^2 \sqrt{2\pi}\frac{\sigma_{en}}{\omega_L} \right]$$

$$= \omega_L^2 \left[ \frac{2\sigma_{en}^2}{\omega_L^2} + \frac{16|\kappa|^2}{\Delta\omega^2} d(t)^2 \right]$$

Thus, the variance is time-indpendent,

$$\Delta x^2(t) = \langle x(t)^2 \rangle - \langle x(t) \rangle^2 = 2\sigma_{en}^2 \tag{30}$$

According to our theoretical analysis, we compare the averaged trajectory $\langle x(t) \rangle$ and the variance $\Delta x(t)^2$ evolve with time during two periods, shown in Fig. S1.

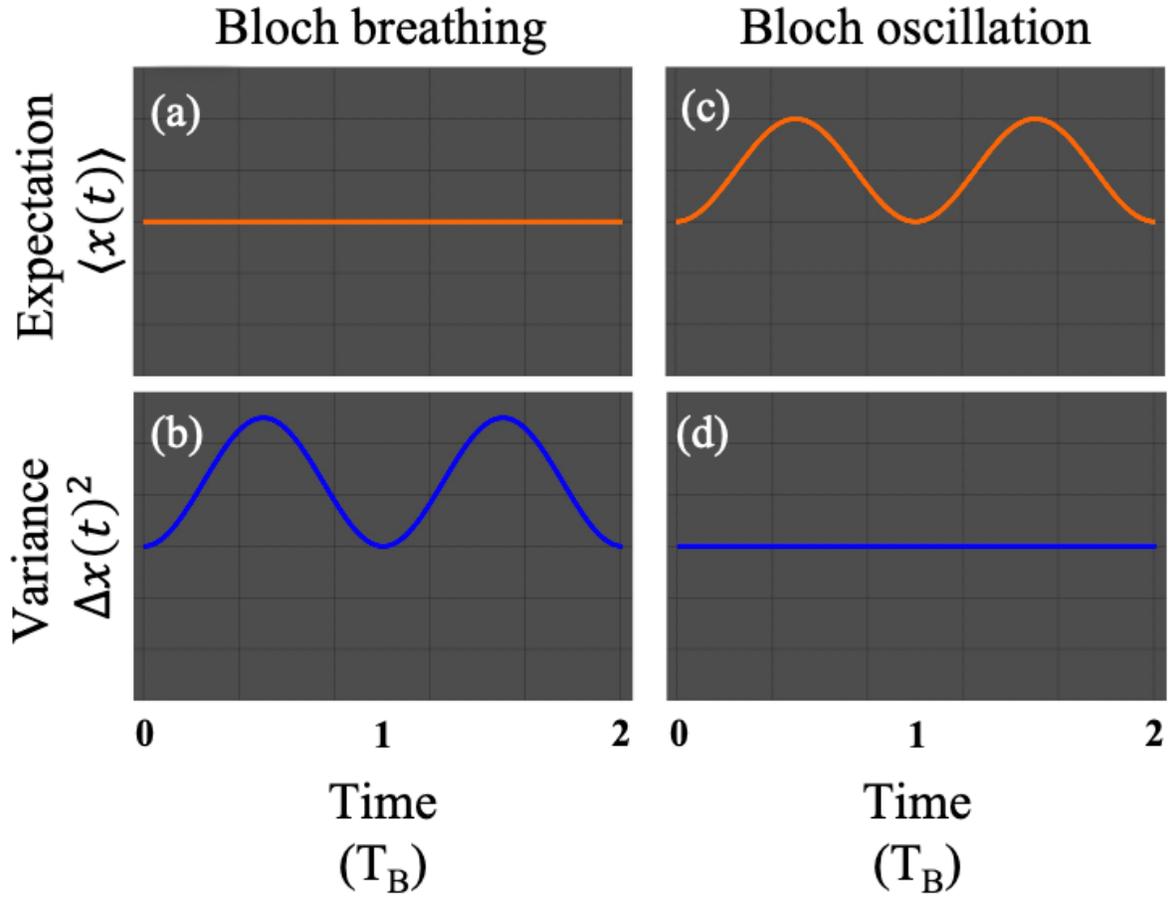

Fig. S1: Comparison between Bloch breathing and Bloch oscillation. For Bloch breathing, the mass-of-the-center trajectory is time-independent (a). However, its variance shows the oscillation feature (b) with the period of $T_B$. On the contrary, the trajectory is time-periodic (c) for Bloch oscillation, but the variance is constant (d). The patterns are calculated with Eqs. (25, 26, 29, 30).

## 5. Algorithm for solving the TDSE of the laser-modulated ultrafast electrons

In order to double-check our result, we start from the initial Schrodinger equation

$$i\hbar\frac{\partial}{\partial t}\psi(z,t) = \widehat{H}\psi(z,t) = \left(\widehat{H}_0 + \widehat{H}_I\right)\psi(z,t) \tag{31}$$

Where the kinetic Hamiltonian of free electron is $\widehat{H}_0 = E_0 + v_0(\hat{p} - p_0) + \frac{(\hat{p}-p_0)^2}{2\gamma^3 m}$, and the interaction Hamiltonian can be written as $\widehat{H}_I = -\frac{eA_0}{\gamma m}\sin(\omega_L t - k_z z + \phi_0) \cdot \hat{p}$ under the Coulomb gauge. Writing the wavefunction as a product of the slow-moving part and the propagation part:

$$\psi(z,t) = \chi(z,t)e^{-\frac{i(E_0 t - p_0 z)t}{\hbar}} \tag{32}$$

Thus, the LHS of Eq. (31) is

$$i\hbar\frac{\partial}{\partial t}\psi(z,t) = e^{-\frac{i(E_0 t - p_0 z)t}{\hbar}}\left(E_0 + i\hbar\frac{\partial}{\partial t}\right)\chi(z,t) \tag{33}$$

And the RHS is

$$\begin{aligned}
&[e^{-\frac{i(E_0 t - p_0 z)t}{\hbar}}\left(E_0 + v_0\big((\hat{p}+p_0)-p_0\big) + \frac{\big((\hat{p}+p_0)-p_0\big)^2}{2\gamma^3 m}\right) \\
&\quad - e^{-\frac{i(E_0 t - p_0 z)t}{\hbar}}\frac{eA_0}{\gamma m}\sin(\omega_L t - k_z z + \phi_0)\cdot(\hat{p}+p_0)]\chi(z,t) \\
&= e^{-\frac{i(E_0 t - p_0 z)t}{\hbar}}\left(E_0 + v_0\hat{p} + \frac{\hat{p}^2}{2\gamma^3 m} - \frac{eA_0}{\gamma m}\sin(\omega_L t - k_z z + \phi_0)\cdot(\hat{p}+p_0)\right)\chi(z,t)
\end{aligned} \tag{34}$$

Thus, we modify the original equation into

$$\begin{aligned}
i\hbar\frac{\partial}{\partial t}\chi(z,t) = &\left(\frac{\hat{p}^2}{2\gamma^3 m} + \left[v_0 - \frac{eA_0}{\gamma m}\sin(\omega_L t - k_z z + \phi_0)\right]\hat{p}\right)\chi(z,t) \\
&- \frac{eA_0 p_0}{\gamma m}\sin(\omega_L t - k_z z + \phi_0)\,\chi(z,t)
\end{aligned} \tag{35}$$

Substitute the momentum operator $\hat{p} = -i\hbar\frac{\partial}{\partial z}$ and $p_0 = \gamma\beta mc$, $A_0 = E_0/\omega_L$

$$i\hbar\frac{\partial}{\partial t}\chi(z,t) = \left(-i\hbar\left[v_0 - \frac{eE_0}{\gamma m\omega_L}\sin(\omega_L t - k_z z + \phi_0)\right]\frac{\partial}{\partial z}\right)\chi(z,t)$$
$$-\left[\frac{\hbar^2}{2\gamma^3 m}\frac{\partial^2}{\partial z^2} + \frac{eE_0\beta c}{\omega_L}\sin(\omega_L t - k_z z + \phi_0)\right]\chi(z,t) \tag{36}$$

Divided by $\hbar c$ on both side

$$i\frac{\partial}{\partial\tau}\chi(z,\tau) = -i\left[\beta - \frac{eE_0}{\gamma mc\omega_L}\sin(k_L\tau - k_z z + \phi_0)\right]\frac{\partial}{\partial z}\chi(z,\tau)$$
$$-\left[\frac{\hbar}{2\gamma^3 mc}\frac{\partial^2}{\partial z^2} + \frac{eE_0\beta}{\hbar\omega_L}\sin(k_L\tau - k_z z + \phi_0)\right]\chi(z,\tau) \tag{37}$$

where $\tau = ct$ and $k_L = \frac{\omega_L}{c}$. All the variables are calculated in the length unit of $\mu m$, and the time of $fs$. Assume the electric field on the grating surface is $10^7\,V/m = 10\,V/\mu m$. Define

$$\alpha_1 = \frac{eE_0}{\gamma mc\omega_L} = 1.77\times 10^{-6} \qquad \alpha_2 = \frac{\hbar}{2\gamma^3 mc} = 4.4\times 10^{-7}\mu m \qquad \alpha_0 = \frac{eE_0\beta}{\hbar\omega_L} = 0.72\,\mu m^{-1}$$

Thus, the simplified equation is

$$i\frac{\partial}{\partial\tau}\chi(z,\tau) = -i[\beta - \alpha_1\sin(k_L\tau - k_z z + \phi_0)]\frac{\partial}{\partial z}\chi(z,\tau)$$
$$-\left[\alpha_2\frac{\partial^2}{\partial z^2} + \alpha_0\sin(k_L\tau - k_z z + \phi_0)\right]\chi(z,\tau) \tag{38}$$

Then we apply the discretization for the spatial parameter $z$ by dividing the spatial domain into $N_z$ parts, with the interval $\delta z = (z_{max} - z_{min})/N_z$

$$\frac{\partial^2}{\partial z^2}\chi(z,t) = \frac{1}{\delta z^2}[\chi(z+\delta z,t) + \chi(z-\delta z,t) - 2\chi(z,t)]$$
$$-i\beta\frac{\partial}{\partial z}\chi(z,t) = -\frac{i\beta}{2\delta z}[\chi(z+\delta z,t) - \chi(z-\delta z,t)]$$
$$i\alpha_1\sin(k_L\tau - k_z z + \phi_0)\frac{\partial}{\partial z} = -\frac{\alpha_1}{\delta z}\cos(k_L\tau - k_z z + \phi_0)\,\chi(z,t)$$
$$+\frac{\alpha_1}{2\delta z}\left[e^{i(k_L\tau - k_z z + \phi_0)}\chi(z+\delta z,t) + e^{-i(k_L\tau - k_z z + \phi_0)}\chi(z-\delta z,t)\right] \tag{39}$$

Thus, the discretized Hamiltonian becomes

$$
i\frac{\partial}{\partial\tau}\chi(z,\tau) = \left[-\frac{\alpha_2}{\delta z^2} - \frac{i\beta}{2\delta z} + \frac{\alpha_1}{2\delta z}e^{i(k_L\tau - k_z z + \phi_0)}\right]\chi(z+\delta z, t)
$$
$$
+ \left[-\frac{\alpha_2}{\delta z^2} + \frac{i\beta}{2\delta z} + \frac{\alpha_1}{2\delta z}e^{-i(k_L\tau - k_z z + \phi_0)}\right]\chi(z-\delta z, t) \qquad (40)
$$
$$
\left[\frac{2\alpha_2}{\delta z^2} - \frac{\alpha_1}{\delta z}\cos(k_L\tau - k_z z + \phi_0) - \alpha_0\sin(k_L\tau - k_z z + \phi_0)\right]\chi(z,t)
$$

In order to compute the time evolution of the wave function $\chi(z,\tau)$ for a given initial state $\chi(z,\tau_0)$, we define the vector wavefunction with $N_z$ components

$$
v(\tau) = \left(\chi(z_1,\tau), \chi(z_1,\tau), \dots, \chi(z_1,\tau)\right)^{\mathrm{T}} \qquad (41)
$$

Then, we apply the discretization for the time interval $\tau - \tau_0$ by equally dividing it into $N$ parts with the subinterval $\Delta\tau = (\tau - \tau_0)/N_t$. We use an implicit Crank-Nicholson integrator to propagate the vector wavefunction from one time step to the next. The formal solution to Eq. (41) could be expressed in terms of the time evolution operator as,

$$
v(\tau + \Delta\tau) = U(\tau + \Delta\tau, \tau)v(\tau) \qquad (42)
$$

where the time evolution operator can be expressed as

$$
U(\tau + \Delta\tau, \tau) = (1 + i\,\Delta\tau H/2)^{-1}(1 - i\,\Delta\tau H/2) \qquad (43)
$$

## 6. Mimicking discrete linear optics with PINEM electrons in synthetic space

This section restricts our discussion at the synchronization condition $\Delta\omega_L = 0$.

**Diffraction management -** According to the dispersion relation Eq. (19), we can derive the group velocity and diffraction coefficient of the electron motion in the PINEM synthetic dimension without detuning ($\Delta\omega_L = 0$):

$$\tilde{v}_g = -2|\kappa|\omega_L \cos(\tilde{k}\omega_L - \phi_0)$$
$$\tilde{D} = \frac{\partial \tilde{v}_g}{\partial \tilde{k}} = 2|\kappa|\omega_L^2 \sin(\tilde{k}\omega_L - \phi_0) \tag{44}$$

When the input electron state is a Gaussian-enveloped PINEM,

$$\psi_p = \frac{1}{\sqrt{2\pi\sigma_p\sigma_{en}}}\exp\left[\frac{(p-p_0)^2}{4\sigma_{en}^2}\right]\sum_{n=-\infty}^{\infty}\exp\left[\frac{(p-p_0-n\,\hbar\omega_L/v_0)^2}{4\sigma_p^2}\right] = \sum_{n=-\infty}^{\infty}\psi_n \tag{45}$$

where $\sigma_{en}$ is the envelope width and $\sigma_E$ is the sideband width. When applying the discrete Fourier transform to the wavefunction: $\psi(\tilde{k}) = \sum_n \psi_n\, e^{in\tilde{k}\omega_L}$, it is easy to find that the synthetic momentum wavefunction is localized around $\tilde{k} = 0$. Thus, we get the picture of the group velocity, and the diffraction coefficient varies as the function of initial phase $\phi_0$, shown in Fig. 5a-5f in the main text.

The spectral evolution of the electron with time is highly dependent on the initial phase $\phi_0$, which is the phase of the modulation laser. When the diffraction coefficient $\tilde{D} = 0$, the spectral width remains constant, and the spectral center manifests linear acceleration (Fig. 5c) and deceleration (Fig. 5e) during the evolution, corresponding to the case $\phi_0 = 0$ (acceleration) and $\phi_0 = \pi$ (deceleration), respectively. Most importantly, at $\tilde{D} = 0$, we can realize the refraction and negative refraction using synthetic PINEM electron; see the discussion in the next section. On the other hand, for the diffraction coefficient $\tilde{D} = \pm 1$, the spectral center is not affected. In contrast, the spectral width is broadened according to the anomalous (Fig. 5d) and normal (Fig. 5f) diffraction.

**Perfect imaging -** With the control over the phase of modulation laser, we can realize the perfect imaging in the PINEM synthetic dimension. The general spectral perfect imaging in the synthetic lattice is shown in Fig. S2. We perform different modulations on different regions of the grating, the single wavepacket experiences the expansion and compression process during the whole time evolution, and the final electron returns to the initial state in the PINEM lattice. This phenomenon is analogous to the perfect imaging of a point light source by real-space "superlens."

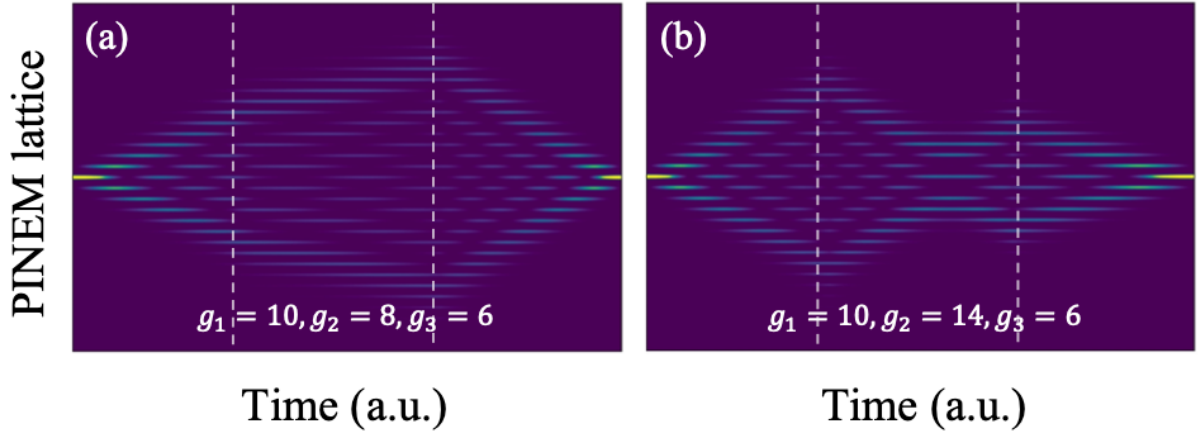

Fig. S2: the general spectral perfect lensing for three-step modulation, whit three sets of coupling constants.

Here we simplify the modulation on a grating as a coupling coefficient $g(t) = \frac{e}{2\hbar\omega} \int_0^t E(t')dt'$ for each step, where the grating length determines the upper limit of integration $t = z/v_e$. The phase modulation can be written as $\hat{M}(g, \phi) = \exp\left[-i|g|\sin\left(\frac{\delta p_L \hat{z}}{\hbar} + \phi\right)\right]$, where $\delta p_L = \frac{\hbar\omega_L}{v_0}$ is the quantum recoil, and $\phi$ is the phase delay. Thus, the final state of the QEW after three-step modulation shown in Fig. S2 can be expressed as

$$\psi_f = \hat{M}(g_3, \phi_3)\hat{M}(g_2, \phi_2)\hat{M}(g_1, \phi_1)\,\psi_i(z) \tag{46}$$

where $\psi_i(p) = \frac{1}{(2\pi\sigma_p^2)^{1/4}} \exp\left(-\frac{p^2}{4\sigma_p^2}\right)$ with $\sigma_p \ll \hbar\omega_L/v_0$.

For the general three-step modulation, express the coupling constant in a vector form: $\boldsymbol{g}_i = (g_i \cos \phi_i, g_1 \sin \phi_i)$, then the condition for the perfect imaging is

$$\boldsymbol{g_1} + \boldsymbol{g_2} + \boldsymbol{g_3} = \boldsymbol{0} \tag{47}$$

Generally, this condition can be extended as $\sum_{i=1}^{N} \boldsymbol{g}_i = 0$ for arbitrary N-step modulation. In Fig. S2, we give the intensity of the three coupling constants, and the phase information can be calculated according to the cosine law.

**The synthetic refraction (the Snell's formula)**, as shown in Fig. S3, is realized in a two-step phase modulation process for an input PINEM electron with a Gaussian envelope. The relative refraction index can be defined as

$$n_{1,2} = \frac{\sin \theta_i}{\sin \theta_r} = \frac{\tilde{v}_g(\phi_1)}{\tilde{v}_g(\phi_2)} = \pm \left| \frac{\kappa_1}{\kappa_2} \right| \tag{48}$$

Thus we can precisely generate both the normal refraction (Fig. S3a) and negative refraction (Fig. S3b) in the PINEM lattice. These phenomena are realized in the acceleration/deceleration regimes, which require the diffraction coefficient to be $\widetilde{D}_{1,2} = 0$. For normal refraction, the group velocity $\tilde{v}_g$ keeps the sign after refraction, thus $\phi_2 - \phi_1 = 0$. While the group velocity $\tilde{v}_g$ changes the sign after negation refraction, thus $\phi_2 - \phi_1 = \pi$. In our simulation, the evolution can be described as

$$\psi(p, t) = e^{-ig_2(t)\sin(\hat{z} + \phi_2)} e^{-ig_1(t)\sin(\hat{z} + \phi_1)} \psi_i(p) \tag{49}$$

The input electron state is $\psi_i = \frac{1}{(2\pi\sigma_p\sigma_{en})^{1/2}} \exp\left(-\frac{p^2}{4\sigma_{en}^2}\right) \sum_n \exp\left[-\frac{(p - n\hbar\omega_L/v_0)^2}{4\sigma_p^2}\right]$, the width of single-mode and the envelope satisfy $\sigma_p \ll \hbar\omega/v_0 \ll \sigma_{en}$.

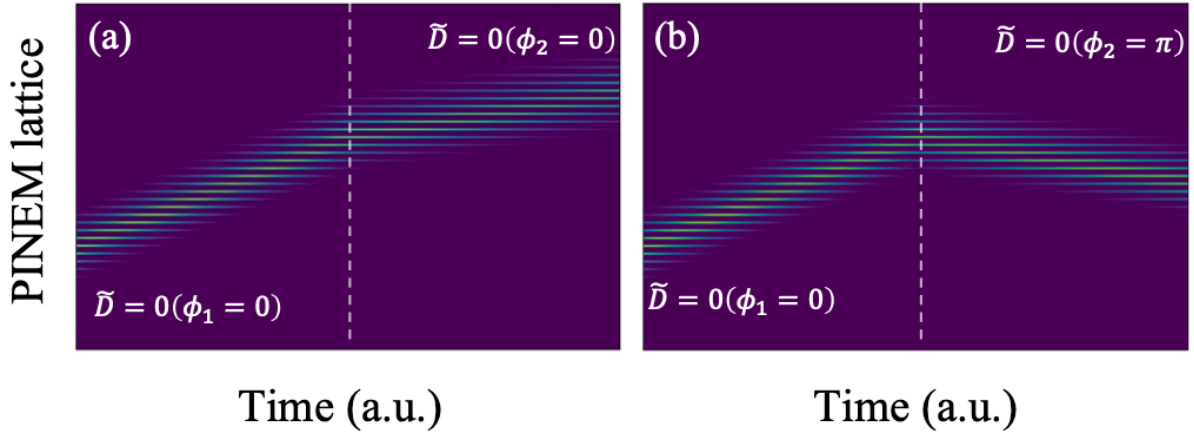

Fig. S3: (a) the normal refraction (b) the negative refraction.

**The discrete Talbot self-imaging effect -** The talbot effect is observed when inputting a periodic array in the PINEM synthetic lattice. We set the periodicity of the PINEM lattice as $T = 10$. Thus the input state has the general form

$$a_n = \sum_m \delta(n - mT) c_m \tag{50}$$

We show two different input state in Fig. S4: (a) $c_m = 1$, (b) $c_m = (-1)^m$ and (c) $c_m = i^m$.

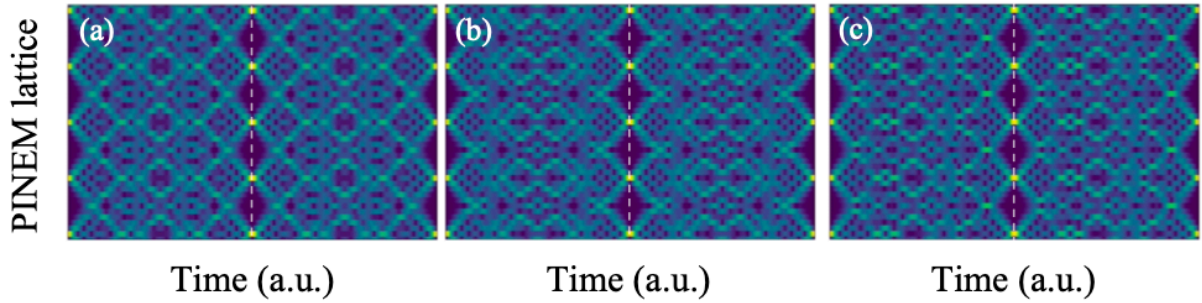

Fig. S4: the Talbot self-imaging with different input PINEM states.